\documentclass[lettersize,journal]{IEEEtran}
\usepackage{amsmath,amsfonts}
\usepackage{bm}
\usepackage{array}
\usepackage{textcomp}
\usepackage{stfloats}
\usepackage{url}
\usepackage{subfigure}
\usepackage{verbatim}
\usepackage{graphicx}
\usepackage{cite,color}
\usepackage{stmaryrd}
\usepackage{algpseudocode}
\usepackage{caption}
\usepackage{hyperref}
\usepackage[ruled,linesnumbered]{algorithm2e}
\usepackage{amsmath,amssymb,amsfonts}
\begin{document}

\title{Data-aided Channel Estimation and Sensing With Sparse Bayesian Learning for AFDM-ISAC System}

\author{Yirui Luo,~\IEEEmembership{Student Member,~IEEE}, Yong Liang Guan,~\IEEEmembership{Senior Member,~IEEE}, Yao Ge,~\IEEEmembership{Member,~IEEE},\\ Yonghong Jiang, Lingsheng Meng, and
David~Gonz\'{a}lez~G.,~\IEEEmembership{Senior Member,~IEEE} 

        \thanks{Yirui Luo, Yong Liang Guan and Lingsheng Meng are with School of Electrical and Electronic Engineering, Nanyang Technological University, Singapore (e-mail: yirui001@e.ntu.edu.sg, eylguan@ntu.edu.sg and meng0071@e.ntu.edu.sg.)}
        \thanks{Yao Ge is with the AUMOVIO-NTU Corporate Lab, Nanyang Technological University, Singapore (e-mail: yao.ge@ntu.edu.sg).}
        \thanks{Yonghong Jiang is with {Electrical Engineering and Computer Science}, University of California, Irvine, USA (e-mail: yonghonj@uci.edu).}
        \thanks{David~Gonz\'{a}lez~G. is with the Wireless Communications Technologies Group, AUMOVIO Germany GmbH (e-mail: david.gonzalez.g@ieee.org).}
        }



\maketitle

\begin{abstract}
Affine frequency division multiplexing (AFDM) has emerged as a promising waveform for next-generation integrated sensing and communication (ISAC) systems. However, it becomes challenging to improve spectral efficiency while simultaneously obtaining accurate channel and sensing-related parameters, particularly in doubly-dispersive channels with fractional delays and fractional Doppler shifts. 
To tackle this challenge, by formulating the channel estimation task as a multiple measurement vectors (MMV) off-grid sparse recovery problem, we propose a data-aided grid-evolution sparse Bayesian learning (D-GESBL) scheme for channel estimation and sensing under a superimposed pilot framework. Specifically, we develop an efficient data-aided iterative receiver, in which reliably decoded data symbols are fed back as additional pseudo-pilot information to assist channel estimation and sensing. To mitigate off-grid mismatch and improve the overall estimation accuracy, we develop a grid evolution procedure that iteratively adjusts the virtual grids in the discrete affine Fourier (DAF) domain according to the estimated off-grid components.
Furthermore, by integrating the generalized approximate message passing (GAMP) algorithm into the proposed SBL framework, we also develop a low-complexity data-aided GAMP–based grid-evolution SBL (D-GAMP-GESBL) algorithm. Finally, the numerical results validate the effectiveness of our proposed schemes and demonstrate their superiority over existing state-of-the-art methods.
\end{abstract}


\section{Introduction}
Emerging applications for the next generation wireless systems, such as unmanned vehicles, large-scale Internet of Things (IoT) and Vehicle-to-Everything (V2X), require both reliable communication and accurate environmental sensing, making the integration of sensing and communication a key trend \cite{Nguyen2022,Z_Zhang2019}. Enabling these dual functions within a unified framework, integrated sensing and communication (ISAC) offers the potential for improved resource efficiency as well as reduced latency and energy consumption, and has therefore attracted considerable attention from both academia and industry \cite{liu}. Among the various ISAC approaches, reusing existing communication signals together with efficient sensing signal processing algorithms is regarded as a practical solution \cite{yirui}. Unfortunately, the orthogonality among the subcarriers of conventional orthogonal frequency division multiplexing (OFDM) is destroyed by the severe Doppler spread in high-mobility scenarios, such as V2X. Consequently, OFDM suffers from strong inter-carrier interference (ICI), leading to substantial performance degradation. This limitation highlights the necessity of developing more robust modulation waveforms that can maintain reliability over doubly dispersive channels in high mobility scenarios.

In recent years, a variety of modulation schemes have been proposed to better accommodate the challenges in high mobility scenarios. One representative example is the orthogonal time frequency space (OTFS) modulation \cite{R_Hadani}, a two dimensional modulation which maps information symbols onto the delay–Doppler (DD) domain rather than the conventional time–frequency (TF) domain. It can fully exploit the inherent delay-Doppler channel diversity, thereby enhancing the performance in the doubly dispersive channels \cite{9508932,10891132}. Besides OTFS, chirp based modulation waveforms with the unique features of the chirp spread spectrum have also been developed for doubly dispersive channels. Orthogonal chirp division multiplexing (OCDM) \cite{OCDM} is one of them and has been shown to outperform OFDM in high-Doppler conditions \cite{OCDM_performance}. However, the achievable diversity and performance of OCDM remains limited due to the fixed chirp rate. The recently proposed affine frequency division multiplexing (AFDM) \cite{AFDM_TWC} is a multi-chirp waveform based on one-dimensional discrete affine Fourier transform (DAFT). By flexibly adjusting the chirp parameters of DAFT according to the channel spread characteristics, AFDM can separate all propagation paths in the DAF domain and allows each transmitted symbol to experience every path coefficient, thereby achieving full diversity in doubly dispersive channels \cite{luoqu_AFDM-SCMA,Luo2025JSGraphAFDM}. Furthermore, existing studies have shown that AFDM can achieve bit error rate (BER) performance comparable to that of OTFS, while requiring lower pilot overhead \cite{AFDM_TWC,AFDM_MIMO} and having lower implementation complexity \cite{K_R_R_Ranasinghe}. Owing to the capability of supporting reliable communication in high-mobility scenarios, and the ability to distinguish multiple targets in the DAF domain, AFDM is considered as a promising candidate for ISAC systems \cite{10858612,AFDM_ICC,11322564}. 

Accurate acquisition of wireless channel parameters is essential for both communication and sensing. 
To achieve this, authors in \cite{bistatic_AFDM} investigated bistatic parameter estimation for static targets by using AFDM, but it does not support Doppler shift estimation. For doubly dispersive channels, authors in \cite{yin_CE_9880774} proposed a threshold-based channel estimation scheme to obtain channel parameters through a mapping table. However, this scheme assumes integer-valued channel delay and Doppler, whereas in practice the finite system bandwidth and AFDM frame duration lead to limited resolution, resulting in fractional delay and Doppler, thus challenging accurate channel parameter estimation. To address this challenge, an approximate maximum likelihood (ML) based approach was introduced in \cite{ML}. 
By refining the searching step size for delay and Doppler shift, the approximate ML approach is capable of estimating fractional delay and Doppler parameters. 
However, the resulting exhaustive two-dimensional grid search imposes a substantial computational burden. To reduce the computational complexity, the authors in \cite{MYIOTJ10963873} proposed a two-stage delay and Doppler shift estimation scheme based on pulse compression. 

To further exploit channel sparsity and enhance parameter estimation performance, compressed sensing–based algorithms have been widely investigated, in which the channel parameter estimation problem is transformed into a sparse recovery problem.
In \cite{4DOMP9891774}, an orthogonal matching pursuit (OMP) based scheme was proposed to estimate the delay–Doppler–angle domain channel of MIMO systems. However, its performance is fundamentally limited by the codebook resolution, as only on-grid components can be recovered. To extract the off-grid components, a Newtonized OMP (NOMP) algorithm based on Newton’s method was introduced in \cite{NOMP9181410}. However, the overall performance remains constrained by the greedy nature of OMP and still exhibits a noticeable gap from the optimal estimation performance. To mitigate the performance limitations of OMP-based methods, sparse Bayesian learning (SBL) framework has emerged as effective alternatives with improved estimation accuracy \cite{SBL11148183}. In conventional SBL, the estimation performance can be enhanced by increasing the resolution of the virtual grid, at the cost of a higher computational complexity due to the enlarged codebook dimension. To strike a balance between performance and complexity, an off-grid SBL (OGSBL) scheme based on first-order Taylor approximation was proposed in \cite{OGSBL9738478}, but failed to properly account for the approximation error introduced by the linearization. In \cite{xiangxiang_11303728}, two dynamic grid schemes were proposed for AFDM channel estimation, but only integer delays were considered. 
Meanwhile, the aforementioned works considered the embedded pilots framework which needs guard symbols to mitigate the mutual interference between pilot and data. However, in short frame   communication systems or under wide Doppler spread, the insertion of guard symbols inevitably leads to a reduction in spectral efficiency (SE).

By superimposing pilots onto data symbols to reduce pilot overhead, superimposed pilot strategy is considered as a feasible approach for improving SE \cite{GAN_10543050}. To minimize the channel estimation error based on linear minimum mean square error (LMMSE) estimator, the authors in \cite{10711268} proposed an effective pilot placement method for AFDM and applied an iterative channel estimation and signal detection scheme. Also, the authors in \cite{10946599} proposed an OMP-based channel estimation scheme to exploit the channel sparsity for AFDM systems with superimposed pilots. To capture the block-sparse structure of the channel and further improve channel estimation performance, the authors in \cite{SP_SBL_SIC_10640141} proposed a novel multiple-frame-based SBL scheme for OTFS systems with a coupled hierarchical prior. Although these methods mitigate pilot–data interference via iterative channel estimation and data detection, the data symbols are consistently treated as interference during channel estimation. Consequently, the valuable information carried by correctly detected data is not exploited, which ultimately limits the achievable performance. The authors in \cite{DA_OTFS_9539066} developed a data-aided scheme based on minimum mean square error (MMSE) channel estimation for OTFS systems. Although this scheme fully exploits the detected data within the iterative process, the assumption of integer-valued delay and Doppler is unrealistic in practice. Moreover, the aforementioned works are restricted to single-antenna settings and thus cannot exploit the spatial domain for performance enhancement and apply the angle information for target sensing.

Against the background, we propose a data-aided grid evolution SBL (D-GESBL) channel estimation and sensing scheme to address the multiple measurement vectors (MMV) sparse recovery problem for AFDM-ISAC systems in this paper. To avoid the need of matrix inversion, we also develop a data-aided generalized approximate message passing (GAMP)–GESBL scheme for computational complexity reduction.
The main contributions of our work are summarized as follows:

\begin{itemize}
\item[1)]
We propose a data-aided channel sensing framework for AFDM-ISAC systems in the presence of fractional delay and fractional Doppler shift by exploiting superimposed pilots. The detected data symbols are iteratively fed back as auxiliary observations to refine the channel estimation. By jointly utilizing the pilot and the progressively improved data decisions, the proposed framework can achieve high-accuracy channel estimation and sensing.
\end{itemize}
 
\begin{itemize}
\item[2)]
To utilize the sparsity of the channel, we formulate the channel parameters estimation task as a MMV off-grid sparse signal recovery problem with the decoded data as the side information for performance enhancement, and propose a D-GESBL scheme for channel estimation and sensing. By employing a dynamic grid-evolution strategy, the virtual grids in the DAF domain are non-uniformly refined according to the estimated on-grid and off-grid components, which significantly mitigates off-grid mismatch and improves the overall channel estimation and sensing performance.
\end{itemize}

\begin{itemize}
\item[3)]
To address the substantial computational complexity introduced by the matrix inversions in the D-GESBL scheme, we further develop a low-complexity data-aided GAMP-GESBL (D-GAMP-GESBL) scheme. Specifically, the GAMP framework replaces the expensive matrix inversion operations with low-complexity iterative message-passing updates, while remaining compatible with the dynamically updated grids. This extension preserves the advantages of grid evolution and data-aided refinement, yet substantially reduces the computational overhead, thereby making the proposed sensing scheme more efficient when the codebook is large.
\end{itemize}
 
\begin{itemize}
\item[4)]
{Simulation results demonstrate the effectiveness of the proposed data-aided scheme, its fast convergence behavior, and the mutual enhancement between channel estimation and data detection. The proposed D-GAMP-GESBL scheme achieves a lower computational complexity than the proposed D-GESBL scheme without significantly performance loss. Moreover, both proposed schemes outperform their corresponding SIC-based counterparts and achieve superior sensing accuracy compared with existing benchmark schemes, while also achieving performance close to the genie bound.}
\end{itemize}

The remainder of this paper is organized as follows. Section II introduces the superimposed pilot framework and the signal model of the AFDM-ISAC system. Section III explains the data-aided SBL strategy and Section IV explains our proposed GESBL scheme for data-aided ISAC. Section V introduces our proposed low complexity D-GAMP-GESBL scheme, followed by the numerical results in Section VI. Section VII concludes the paper.

Notation: Unless otherwise stated, we use the lowercase bold letter and uppercase bold letter denote vectors and matrices, respectively. $x_i$, $X_{i,j}$ and $\mathbf{X}_i$ represent the $i$-th element of vector ${\mathbf{x}}$, $\left(i,j\right)$-th element of the matrix ${\mathbf{X}}$ and the $i$-th column of the matrix ${\mathbf{X}}$, respectively. $\text{diag}\left(  {\mathbf{x}}  \right)$ returns a diagonal matrix whose diagonal elements are vector ${\mathbf{x}}$, $\text{diag}\left(  {\mathbf{X}}  \right)$ stands for a vector consisting of the diagonals of the matrix ${\mathbf{X}}$ and $\text{vec}\left(\mathbf{X}\right)$ returns a column vector formed by stacking all columns of matrix $\mathbf{X}$. $\operatorname{Re} \left\{ \cdot \right\}$ denotes the real part of the input. A random complex-valued vector ${\mathbf{x}} \in {\mathbb{C}^{N\times 1}}$ which follows the complex Gaussian distribution is defined as $\mathcal{C}\mathcal{N}\left( {{\mathbf{x}};{\bm{\mu }},{\bm{\Sigma }}} \right) = {\left( {{\pi ^N}\left| {\bm{\Sigma }} \right|} \right)^{ - 1}}\exp \left\{ { - {{\left( {{\mathbf{x}} - {\bm{\mu }}} \right)}^H}{{\bm{\Sigma }}^{ - 1}}\left( {{\mathbf{x}} - {\bm{\mu }}} \right)} \right\}$. $\mathbb{E}\{\cdot\}$ stands for the expectation. ${\left(  \cdot  \right)^{ - 1}}$, ${\left(  \cdot  \right)^{ *}}$, ${\left(\cdot\right)^{T}}$, and ${\left(\cdot\right)^{H}}$ are the inverse, conjugate, transpose, and conjugate transpose, respectively. 

\section{System Model}
\begin{figure*}
\centering
\includegraphics[width=18cm]{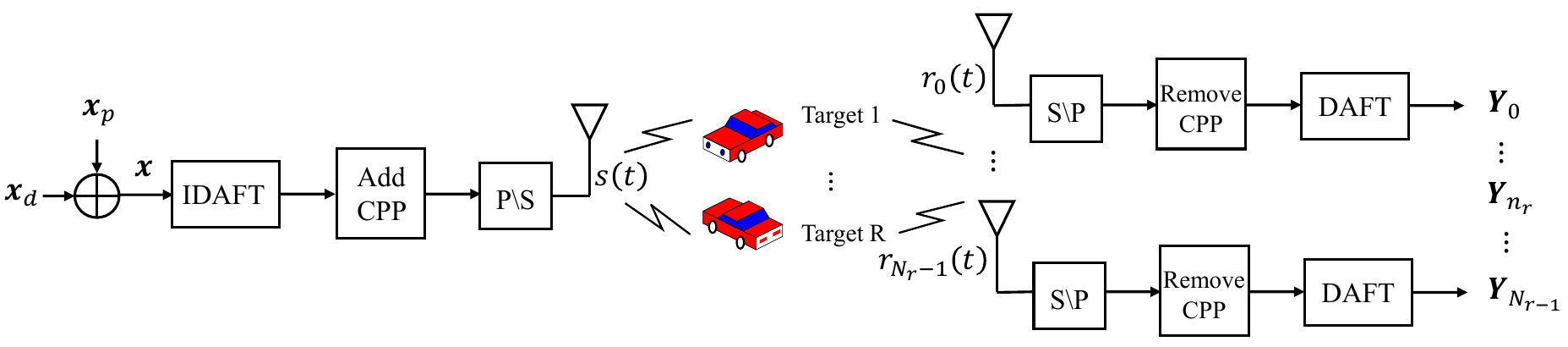}
\caption{System Model.}
\label{system model}
\end{figure*}

As illustrated in Fig. \ref{system model}, we consider a sensing integrated uplink AFDM system aided by superimposed pilots, operating over a channel with a carrier frequency $f_c$. Specifically, the single-antenna UE serves as the ISAC transmitter and sends an AFDM-ISAC waveform with superimposed pilots to the base station (BS), which functions as the ISAC receiver and is equipped with a uniform linear array (ULA) of $N_r$ antennas. After acquiring the received signal, the BS simultaneously recovers the communication data and senses the channel and surrounding target parameters. Furthermore, the detected data symbols are exploited in a data-aided manner to refine channel and target estimation sensing accuracy.

\subsection{AFDM Transmit Signal Design}
\begin{figure}[!t]
\centering
\includegraphics[width=3.5in]{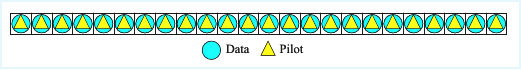}
\caption{Superimposed pilot framework in the DAF domain.}
\label{SP}
\end{figure}

Consider an AFDM symbol with a bandwidth $BW=N\Delta f$ and a duration of $T$, where $N$ is the number of the chirp subcarriers and $\Delta f=1/T$ is the chirp subcarrier spacing. Let $\bm{x}_d$ be an $N \times 1$ QAM symbol vector in the DAF domain. As shown in Fig. \ref{SP}
, the pilot symbol $x_p\left[m\right]$ on the $m$-th chirp subcarrier is superimposed onto the data symbol ${x_d}\left[m\right]$ in the DAF domain as ${x}\left[m\right]={x_p}\left[m\right]+{x_d}\left[m\right]$, which can also be expressed in a vector form as
\begin{equation}
    \bm{x} = \bm{x}_p+\bm{x}_d,
\end{equation}
where $\bm{x} \in \mathbb{C}^{N\times 1}$ is the transmitted symbols in the DAF domain. The pilot and data symbols are modeled as zero-mean independent and identically distributed (iid) random variables with variances $\mathbb{E}\{|x_p\left[m\right]|^2\}=\sigma_p^2$ and $\mathbb{E}\{|x_d\left[m\right]|^2\}=\sigma_d^2$, respectively. A power normalization is imposed such that the transmitted symbol satisfies $\mathbb{E}\{|x\left[m\right]|^2\}=\sigma_p^2+\sigma_d^2=1$ \cite{SP_SBL_SIC_10640141}.

To obtain the time-domain signal, an $N$-point inverse DAFT (IDAFT) is applied to $\bm{x}$, resulting in
\begin{equation}
    s\left[ n \right] = \frac{1}{{\sqrt N }}\sum\limits_{m = 0}^{N - 1} {x\left[ m \right]e^ {{j2\pi \left( {{c_1}{n^2} + {c_2}{m^2} + nm/N} \right)} }},\label{s[n]}
\end{equation}
for $n=0,\dots,N-1$. $c_1$ and $c_2$ are the AFDM chirp parameters, whose proper selection ensures full diversity. In matrix notation, \eqref{s[n]} can be compactly written as ${\bm{s}} = {\boldsymbol{\Lambda }}_{{c_1}}^H\mathbf{F}^H{\boldsymbol{\Lambda }}_{{c_2}}^H{\bm{x}}=\mathbf{A}^H\bm{x}$, where $\mathbf{F}$ denotes the normalized $N$-point discrete Fourier transform (DFT) matrix and ${{\mathbf{\Lambda }}_{c_p}} = {\text{diag}}\left( {{e^{ - j2\pi c_p{n}^2}}, p=1,2} \right)$.
To maintain circularity in multipath channels, a chirp-periodic prefix (CPP) of length $N_\text{cpp}$ is appended prior to the transmitted signal $\mathbf{s}$, defined as 
\begin{equation}
    s[n] = s\left[ {N + n} \right]{{\text{e}}^{ - j2\pi {c_1}\left( {{N^2} + 2Nn} \right)}},{\text{     }}n =  - {N_\text{cpp}},..., - 1,
\end{equation}
with $N_\text{cpp}$ chosen to be an integer no smaller than the maximum channel delay taps. The transmitted AFDM signal can be described in continuous time as
\begin{equation}
    s\left( t \right) = \frac{1}{{\sqrt T }}\sum\limits_{m = 0}^{N - 1} {x\left[ m \right]} {{\text{e}}^{j2\pi \left( c_2m^2+\Phi_m\left(t\right)\right)}},{\text{  }}0 \leqslant t < T,
\end{equation}
where the phase term $\Phi_m(t)$ is defined piecewise over the interval partition $\{t_{m,q}\}_{q=0,...,2Nc_1}$ of $[0, T)$ as $\Phi_m\left(t\right)=\frac{{{c_1}}}{{T_s^2}}{t^2} + \frac{m}{{N{T_s}}}t-\frac{q}{T_s}t$ for $t\in\left[t_{m,q},t_{m,q+1}\right)$, with $t_{m,0}=0$ and $t_{m,q}=\frac{N-m}{2Nc_1}T_s+\frac{q-1}{2c_1}T_s$ for $q>0$. $T_s=T/N$ denotes the sampling interval.
\subsection{Channel Model}
The time domain AFDM signal $s\left(t\right)$ is propagated through a doubly dispersive sensing channel containing $R$ targets, where each target is assumed to generate a single propagation path. Let $\gamma_i$, $L_i$, $v_i$, and $\theta_i$ denote the scattering coefficient, total range, relative velocity and angle of arrival (AoA) of target $i$, respectively. The corresponding delay and Doppler shift are given by $\tau_{i}=\frac{L_i}{c}$ and $f_{d,i}=\frac{v_i f_c}{c}$, where $c$ is the speed of light. Thus, the matrix impulse response for the time-varying sensing channel can be modeled as
\begin{equation}
    {\mathbf{H}}\left( {t,\tau } \right) = {\sum\limits_{i = 0}^{R}  {{\gamma_i}{\mathbf{a}}_R\left( {{\theta _i}}\right)}}\delta \left( {\tau  - {\tau _i}} \right){e^{j2\pi {f_{d,i}}t}},\label{channel}
\end{equation}
where $i=0$ corresponds to the direct line-of-sight path between the UE and the BS. The receive array steering vector {${\mathbf{a}_R}\left( \theta_i  \right)\in \mathbb{C}^{1\times N_r}$} for a ULA with half-wavelength spacing can be expressed as 
\begin{equation}
      {\mathbf{a}_R}\left( \theta_i  \right) = {\left[1,{{e^{ - j\pi \sin \theta_i }}},...,{{e^{ - j\pi\left( {N_r - 1} \right)\sin \theta_i }}} \right]}.\label{AOA_steer}
 \end{equation}

\subsection{Received signals}
{The time-domain AFDM signal received at the $n_r$-th antenna of BS can be represented as
\begin{equation}
\begin{split}
    {r}_{n_r}\left( t \right) =&\sum\limits_{i = 0}^{R} {{\gamma_i} \left[{\mathbf{a}}_R\left( {\theta _i}\right)\right]_{n_r}} {e^{j2\pi {f_{d,i}}t}}{s}\left( {t-\tau_{i}} \right)  + {w}_{n_r}\left( t \right), 
\end{split} 
\end{equation}
 where $\left[{\mathbf{a}}_R\left( {\theta _i}\right)\right]_{n_r}$ and  ${w}_{n_r}\left(t\right)$ denotes the $n_r$-th element of ${\mathbf{a}}_R\left( {\theta _i}\right)$ and the additive Gaussian noise observed at the $n_r$-th antenna in the time domain, respectively, with $n_r=0,...,N_r-1$. 
Upon sampling the continuous-time signal with interval $T_s$, then following serial to parallel conversion, CPP removal and DAFT, the signal received by the $n_r$-th antenna $\mathbf{Y}_{n_r}\in \mathbb{C}^{N\times 1}$ in the DAF domain can be expressed as
\begin{equation}
\begin{split}
    {\mathbf{Y}_{n_r}} &= \sum\limits_{i = 0}^{R} {{\gamma_i}e^{-j\pi n_r\sin \theta_i}}\mathbf{A}{\mathbf{\Gamma }}_{CP{P_i}}{{\mathbf{\Delta }}_{{\nu_i}}}{{\mathbf{\Pi }}^{\eta_i}}\mathbf{A}^H\bm{x}+ {\mathbf{W}_{n_r}}\\
    &=\mathbf{H}_{eff,n_r}\bm{x}+ {\mathbf{W}_{n_r}}
    ,\label{DAF-receive}
\end{split}
\end{equation}
where ${\mathbf{W}_{n_r}}$ is the corresponding noise received by the $n_r$-th antenna in the DAF domain. ${{\boldsymbol{\Gamma }}_{CP{P_i}}}$ is a diagonal matrix with the expression of
 \begin{equation}\label{CPP}
\begin{split}
{{\boldsymbol{\Gamma }}_{CP{P_i}}} = {\text{diag}}\left( {\left\{ {\begin{array}{*{20}{c}}
  {{{{e}}^{ - j2\pi {c_1}\left( {{N^2} - 2N\left( {{\ell_i} - n} \right)} \right)}}}&{n < {\ell_i}} \\ 
  1&{n \geqslant {\ell_i}} 
\end{array}} \right.} \right).
\end{split}
\end{equation}
From \eqref{CPP}, it can be observed that when $N$ is even and $2Nc_1$ is an integer, $\boldsymbol{\Gamma}_{CPP_i}=\mathbf{I}$. Therefore, in the subsequent derivations, we set $\boldsymbol{\Gamma}_{CPP_i}=\mathbf{I}$.
The Doppler shift related matrix ${{\mathbf{\Delta }}_{{\nu_i}}}$ is given by ${{\mathbf{\Delta }}_{{\nu_i}}} = {\text{diag}}({e^{ - j2\pi {{\left( {0:{N} - 1} \right){\nu_i}} /{N}}}})$ and the delay related matrix ${\mathbf{\Pi }}^{\eta_i}$ is defined as ${\mathbf{\Pi }}^{\eta_i}={{\mathbf{F}}^H}{\text{diag}}({e^{ - j2\pi {\left( {0:{N} - 1} \right) \eta_i}/ {N}}}){\mathbf{F}}$, where $\eta_i=\tau_i/T_s$ and $\nu_i=Nf_{d,i}T_s$ denote the normalized
delay and Doppler shift associated with the $i$-th target, respectively. Each of these parameters can be decomposed into an integer and a fractional component, expressed as $\eta_i=\ell_i+\iota_i$ and $\nu_i=k_i+\kappa_i$. $\ell_i \in \left(0,\ell_{\max}\right]$ and $\iota_i\in \left( { - \frac{1}{2}} \right.,\left. {\frac{1}{2}} \right]$ denote the integer and fractional normalized delays, while $k_i \in \left[-k_{\max},k_{\max}\right]$ and $\kappa_i\in \left( { - \frac{1}{2}} \right.,\left. {\frac{1}{2}} \right]$ represent the corresponding integer and fractional normalized Doppler components. Here, $\ell_{\max}$ and $k_{\max}$ specify the maximum integer normalized delay and Doppler shift in the ISAC system, respectively. $c_1$ is set to be $c_1=\frac{{2\left( {{k _{\max }} + {k_v}} \right) + 1}}{{2N}}$ to preclude collisions of distinct non-zero entries\cite{AFDM_TWC}, where $k_v$ is a non-negative design parameter introduced to suppress the interference caused by fractional Doppler shifts. Furthermore, to ensure full diversity, the system parameters must satisfy
\begin{equation}
    2\left({k_{\max }}+k_v\right) + {\ell_{\max }} + 2\left({k_{\max }}+k_v\right){\ell_{\max }} < N.
\end{equation}}

 By combining the received signals from all antennas, the overall received signal $\mathbf{Y}\in \mathbb{C}^{N\times N_r}$ in the DAF domain can be expressed in the matrix form of
\begin{equation}
\begin{split}
    {\mathbf{Y}} &=\sum\limits_{i = 0}^R {\mathbf{\Xi  }\left(\nu_i,\eta_i\right)\left( {\bm{x}_p + \bm{x}_d} \right)\left( {{\gamma_i}\mathbf{a}_R\left( {{\theta _i}} \right)} \right)} 
    + {\mathbf{W}}\\
    &=\left( {\mathbf{\Psi}_p + \mathbf{\Psi}_d} \right)\mathbf{H}+ {\mathbf{W}}
    ,\label{DAF-receive}
\end{split}
\end{equation}
where 
\begin{equation}
    \mathbf{\Psi}_p=\left[\mathbf{\Xi }\left(\nu_0,\eta_0\right)\bm{x}_p,\mathbf{\Xi}\left(\nu_1,\eta_1\right)\bm{x}_p,...,\mathbf{\Xi }\left(\nu_R,\eta_R\right)\bm{x}_p\right],\label{phi_p}
\end{equation}
\begin{equation}
    \mathbf{\Psi}_d=\left[\mathbf{\Xi }\left(\nu_0,\eta_0\right)\bm{x}_d,\mathbf{\Xi }\left(\nu_1,\eta_1\right)\bm{x}_d,...,\mathbf{\Xi}\left(\nu_R,\eta_R\right)\bm{x}_d\right],\label{phi_d}
\end{equation}
with $\mathbf{\Xi }\left(\nu_i,\eta_i\right)=\mathbf{A}{{\mathbf{\Delta }}_{{\nu_i}}}{{\mathbf{\Pi }}^{\eta_i}}\mathbf{A}^H$. $\mathbf{H}\in\mathbb{C}^{(R+1) \times N_r}$ contains the information of scattering coefficient and AoA, and its $(i,n_r)$-th element has the expression of 
\begin{equation}
    {H}_{i,n_r}={\gamma_i}e^{-j\pi n_r\sin \theta_i}.
\end{equation}

It can be observed from \eqref{DAF-receive} that mutual interference exists between the data and pilot signals. To mitigate this interference, the existing AFDM superimposed-pilot frameworks \cite{10711268} \cite{10946599}  adopt a successive interference cancellation (SIC) approach. First, the data symbols are treated as interference to obtain an initial channel estimation. Then, based on this preliminary estimation, a coarse data detection is performed, and the detected symbols are cancelled from the received signal to suppress data-induced interference, yielding a more refined estimation. By iteratively alternating between data detection and the interference-reduced channel estimation, the channel sensing and communication performance can be gradually improved. However, since the data symbols are always treated as interference and removed in this process, SIC-based methods fail to exploit the potentially valuable information contained in correctly detected data. To overcome this limitation, a data-aided strategy is developed in this paper to leverage reliably decoded data symbols in the codebook to further assist channel estimation and target parameter sensing, yielding superior performance as long as data detection is relatively accurate. The subsequent sections present a detailed description of our proposed data-aided schemes.

\section{Proposed Data-aided SBL Framework}
\begin{figure}
\centering
\includegraphics[width=2.5in]{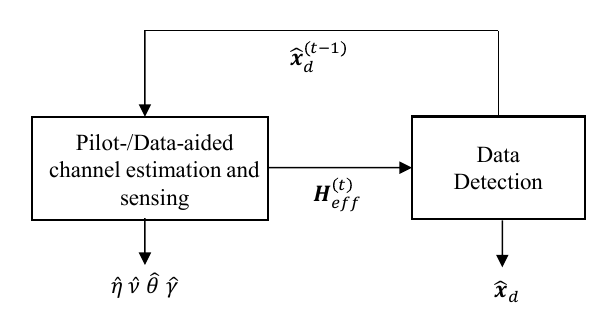}
\caption{{Flow chart of data-aided channel estimation and sensing.}}
\label{chat}
\end{figure}

In this section, we propose a data-aided SBL framework for channel estimation and target parameter sensing, as shown in Fig. \ref{chat}. The key point lies in the iterative exchange of information between channel estimation and data detection.
 \subsection{Data-aided SBL Receiver Framework}
\begin{algorithm}
\caption{Proposed data-aided SBL receiver framework for AFDM-ISAC}\label{data-aided framework}
\KwIn{${\mathbf{Y}}$, ${\ell_{\max }}$, ${k_{\max }}$, maximum iteration $T_{\max }$, the convergence tolerance $\varepsilon_t$}
\textbf{Initialization}:
$t=1$, $\bm{x}_d^{(0)}=\mathbf{0}$\;
\textbf{Repeat}\\
Update $\bm{\hat{x}}^{\left(t\right)}=\bm{x}_p+\bm{\hat x}_d^{\left(t-1\right)}$\;
Update  $\hat{\mathbf{\Phi }}^{\left(t\right)}\left( \bm{\bar k },\bm{\bar \ell } \right)$ according to \eqref{PHI_t}\;
\textbf{Choose \textbf{GESBL} or \textbf{GAMP-GESBL} for channel estimation}:\\
\quad \textbf{- GESBL:} Obtain channel related parameters according to \textbf{Algorithm \ref{GE-SBL}}\;
\quad \textbf{- GAMP-GESBL:} Obtain channel related parameters according to \textbf{Algorithm \ref{GAMP-GE-SBL}}\;
Calculate ${\mathbf{\hat H}}_{eff}^{(t)}$ according to \eqref{H_eff_est}\;
Obtain $\text{vec}\left({\mathbf{Y}}_d^{(t)}\right)$ according to \eqref{Y_d^t}\;
Estimate data symbols $\bm{\hat x}_d^{(t)}$ according to \eqref{data detection}\;
Replace $\bm{\hat x}_d^{(t)}$ by Mod[Demod($\bm{\hat x}_d^{(t)}$)]\;
$t\leftarrow t+1$\;
\textbf{Until} $ \frac{{\left\| {{{\bm{\hat x}_d}^{\left( {t } \right)}} - {{\bm{\hat x}_d}^{\left( t-1 \right)}}} \right\|^2}}{{\left\| {{{\bm{\hat x}_d}^{\left( t-1 \right)}}} \right\|^2}} < \varepsilon_t $ or $t=T_{max}$\\
Calculate ranges and velocities of targets according to the indexes of the $\hat r \in \mathcal{\hat R}$ and estimate the corresponding AoAs according to \eqref{AoA_est}\;
\KwOut{$\mathbf{H}_{eff}$, $\bm{x}_d$, ranges, velocities and AoAs of the targets}
\end{algorithm}

In the first iteration, only the pilot symbols are known, and the data symbols are treated as interference to obtain a coarse estimate of the channel parameters. The coarsely estimated channel parameters are then used to construct the effective channel matrix for data detection. By jointly exploiting the detected data and the superimposed pilots, the channel estimation accuracy is improved, which in turn further enhances the data detection performance. Specifically, with the detected data in the $(t-1)$-th iteration $\bm{\hat{x}_d}^{\left(t-1\right)}$, the received AFDM sensing signal in the $t$-th iteration can be rewritten as
\begin{equation}\label{Y=hatPsiH}
 \begin{split}
    {\mathbf{Y}} &=\left( {\mathbf{\Psi}_p + \mathbf{\hat \Psi}_d^{\left(t-1\right)}} \right)\mathbf{H}+\left(\mathbf{\Psi}_d-\mathbf{\hat \Psi}_d^{\left(t-1\right)}\right)\mathbf{H}+ {\mathbf{W}}\\
    &={\mathbf{\hat \Psi}}^{\left(t\right)}\mathbf{H}+ {\mathbf{\bar{W}}^{\left(t\right)}},
\end{split}
\end{equation}
where $\mathbf{\hat \Psi}^{\left(t\right)}={\mathbf{\Psi}_p + \mathbf{\hat \Psi}_d^{\left(t-1\right)}}$, $\mathbf{\hat \Psi}_d^{\left(0\right)}=\mathbf{0}$ and $\mathbf{\bar{W}}^{\left(t\right)}=\left(\mathbf{\Psi}_d-\mathbf{\hat \Psi}_d^{\left(t-1\right)}\right)\mathbf{H}+ {\mathbf{W}}$ represent the corresponding detection error with noise. Similar to \eqref{phi_p} and \eqref{phi_d}, $ \mathbf{\hat \Psi}_d^{\left(t-1\right)}$ and $ \mathbf{\hat \Psi}^{\left(t\right)}$ have the expression of 
\begin{equation}
\begin{split}
    &\mathbf{\hat \Psi}_d^{\left(t-1\right)}=\\
    &\left[\mathbf{\Xi}\left(\nu_0,\eta_0\right)\bm{\hat x}_d^{\left(t-1\right)},\mathbf{\Xi }\left(\nu_1,\eta_1\right)\bm{\hat x}_d^{\left(t-1\right)},...,\mathbf{\Xi }\left(\nu_R,\eta_R\right)\bm{\hat x}_d^{\left(t-1\right)}\right],
\end{split}
\end{equation}
\begin{equation}
    \mathbf{\hat \Psi}^{\left(t\right)}=\left[\mathbf{\Xi }\left(\nu_0,\eta_0\right)\bm{\hat x}^{\left(t\right)},\mathbf{\Xi }\left(\nu_1,\eta_1\right)\bm{\hat x}^{\left(t\right)},...,\bm{\Xi }\left(\nu_R,\eta_R\right)\bm{\hat x}^{\left(t\right)}\right],\label{hatPsi}
\end{equation}
where $\bm{\hat{x}}^{\left(t\right)}=\bm{x}_p+\bm{\hat x}_d^{\left(t-1\right)}$.

Based on \eqref{hatPsi}, $\mathbf{\hat \Psi}^{\left(t\right)}$ is constructed using the unknown parameters $\nu_i$, $\eta_i$ and $R$. In this situation, both $\mathbf{\hat \Psi}^{\left(t\right)}$ and $\mathbf{H}$, along with their dimensions are unknown, making it infeasible to directly estimate the channel parameters from \eqref{Y=hatPsiH}. To address this challenge, the sensing problem is reformulated as a MMV sparse signal recovery problem through the construction of virtual sampling grids.
{
We first perform a virtual grid sampling along the Doppler and delay dimension in the range of $\left[-k_{max},k_{max}\right]$ and $\left[0,\ell_{max}\right]$, respectively, with Doppler grid resolution $r_\nu$ and delay grid resolution $r_\tau$. Then, ${K_\nu } = \frac{{2{k _{\max }}}}{{{r_k}}} + 1$ and ${L_\tau } = \frac{{{\ell_{\max }} + 1}}{{{r_\tau }}}$ denote the virtual sampling grid sizes along Doppler shift and delay dimension, respectively. To construct the two-dimensional delay-Doppler virtual grids, each Doppler grid point is paired with each delay grid point, resulting in $L_\tau K_\nu$ Doppler and delay grid pairs. For compact representation, these two-dimensional grid pairs are vectorized into two vectors, represented by ${\bm{\bar k}}$ and ${\bm{\bar \ell}}$, which stores the Doppler shift and delay grid values associated with all vectorized delay-Doppler grid pairs, respectively.
Specifically, the virtual grids are expressed as ${\bm{\bar k}} = {\left[ {{{\bar k}_0},{{\bar k}_1},...,{{\bar k}_{{L_\tau }{K_\nu }-1}}} \right]^T} \in {\mathbb{R}^{{L_\tau }{K_\nu }\times 1}}$ and ${\bar{\boldsymbol{\ell}}} = {\left[ {{{\bar \ell}_0},{{\bar \ell}_1},...,{{\bar \ell}_{{L_\tau }{K_\nu }-1}}} \right]^T} \in {\mathbb{R}^{{L_\tau }{K_\nu }\times 1}}$, where ${{\bar k}_j} = k'{r_k} - {k _{\max }}$ and ${{\bar \ell}_j} = \ell'{r_\tau }$, with $j = \ell'{K_\nu } + k'$ for $k' \in {\left\{0,...,{{K_\nu }-1}\right\}}$, $\ell' \in {\left\{0,...,{{L_\tau }-1}\right\}}$ and $j \in {\left\{0,...,{{L_\tau }{K_\nu }-1}\right\}}$.
} Then, the equation \eqref{Y=hatPsiH} can be reformulated as 
\begin{equation}\label{y=Hx}
    {\mathbf{Y}} = {\mathbf{\hat {\Phi} }}^{\left(t\right)}\left(\bm{\bar k },\bm{\bar \ell } \right){\mathbf{\bar H}} + {\mathbf{\bar W}}^{\left(t\right)},
\end{equation}
where the measurement matrix $\hat{\mathbf{\Phi }}^{\left(t\right)}\left(\bm{\bar k },\bm{\bar \ell }\right) \in \mathbb{C}^{N \times {L_\tau }{K_\nu }}$ can be expressed as
\begin{equation}
\begin{split}
    \hat{\mathbf{\Phi }}^{\left(t\right)}\left( \bm{\bar k },\bm{\bar \ell } \right) =\left[\mathbf{\Xi }\left( {\bar k}_0,{\bar \ell}_0 \right)\bm{\hat x}^{\left(t\right)},...,\mathbf{\Xi }\left(  {\bar k}_{L_\tau K_\nu-1},{\bar \ell}_{L_\tau K_\nu-1} \right)\bm{\hat x}^{\left(t\right)}\right].\label{PHI_t}
\end{split} 
\end{equation}
It's worth noting that the unknown matrix ${\mathbf{\bar H}}\in {\mathbb{C}^{{L_\tau }{K_\nu }\times N_r}}$ possesses a common row-sparse structure, meaning that all of its columns are sparse and share an identical support. 

By reformulating \eqref{Y=hatPsiH} into a MMV sparse recovery formulation in \eqref{y=Hx}, we develop a D-GESBL scheme that adaptively refines the grids for channel estimation and target sensing. After obtaining the channel related parameters, an effective channel matrix $\mathbf{\hat H}^{(t)}_{eff,n_r}$ is constructed for each receive antenna ($n_r=0,...,N_r-1$) and can be used for data detection in the $t$-th data-aided iteration. Nevertheless, the required matrix inversions lead to considerable computational complexity. Accordingly, a low complexity GAMP-GESBL scheme for data-aided ISAC is proposed in the following section.

The DAF-domain symbols then can be decoded by removing the pilot interference from the received signal matrix $\mathbf{Y}$ as
\begin{equation}\label{Y_d^t}
\text{vec}\left({\mathbf{Y}}_d^{(t)} \right)= \text{vec}\left({\mathbf{Y}} \right)- {\mathbf{\hat H}}_{eff}^{(t)}{\bm{x}}_p = {\mathbf{\hat H}}_{eff}^{(t)}{\bm{x}}_d + \text{vec}\left({\mathbf{W}}_e^{(t)}\right),
\end{equation}
where ${{\bf{\hat H}}_{eff}^{\left( t \right)}}\in \mathbb{C}^{NN_r\times N}={\left[ {{\bf{\hat H}}_{eff,1}^{(t)T},{\bf{\hat H}}_{eff,2}^{(t)T},...,{\bf{\hat H}}_{eff,{N_r}}^{(t)T}} \right]^{T}}$ and $\text{vec}\left({\mathbf{W}}_e^{(t)}\right) = \text{vec}\left({\mathbf{W}}\right) + \left({\mathbf{H}}_{eff} - {\mathbf{\hat H}}_{eff}^{(t)}\right)\bm{x}_p$ represents the equivalent noise including the channel estimation error. In this paper, we mainly focus on channel sensing scheme design and simply apply LMMSE detector for data detection\footnotemark{}.
\footnotetext{While advanced data detection approaches such as the orthogonal approximate message passing (OAMP) detector \cite{9354639} may yield better BER performance, a detailed study of data detection techniques is beyond the scope of the present paper.}
Accordingly, the DAF-domain data symbols can be recovered by
\begin{equation}
{\bm{x}}_d^{\left( t \right)} = {\left( {{{\left( {{\bf{\hat H}}_{eff}^{\left( t \right)}} \right)}^H}{\bf{\hat H}}_{eff}^{\left( t \right)} + {\sigma_d ^2}{\bf{I}}} \right)^{ - 1}}{\left( {{\bf{\hat H}}_{eff}^{\left( t \right)}} \right)^H}{\rm{vec}}({\bf{Y}}_d^{\left( t \right)}),\label{data detection}
\end{equation}
where $\sigma_d^2$ is the receive noise power for data detection. Then, we demodulate the estimated symbols into the corresponding information bits and then modulate them to symbols, thereby further improving the accuracy of data estimation during the iterative process \cite{GAN_10543050}. Accordingly, the data estimated at the $t$-th iteration is replaced by Mod[Demod($\bm{\hat x}_d^{(t)}$)]. With the updated data vector, the channel estimation process is further refined. A brief explanation of data-aided receiver framework is summarized in \textbf{Algorithm \ref{data-aided framework}}. More details are provided in the following sections.

\subsection{Hierarchical Prior for Sparse Bayesian Learning}
According to \eqref{y=Hx} and the row sparsity of $\mathbf{\bar H}$, we develop the sparse Bayesian model by employing a hierarchical prior that is commonly used in SBL \cite{Tipping2001SBL}. It is assumed that the elements of $\mathbf{\bar H}$ follow a zero-mean Gaussian prior distribution, where each row is associated with an independent variance parameter $\delta_j$ and can be expressed as
\begin{equation}\label{h_prior}
    p\left( {{\mathbf{\bar H}}|{\bm{\delta }}} \right) = \prod_{n_r=0}^{N_r-1} \mathcal{C}\mathcal{N}\left( {{\mathbf{\bar H}_{n_r}};{\mathbf{0}},{\bm{\Delta }}} \right),
\end{equation}
where ${\mathbf{\bar H}_{n_r}}$ is the $n_r$-th column of ${\mathbf{\bar H}}$ and ${\bm{\Delta }} = {\text{diag}}\left( {\bm{\delta }} \right)$. The vector ${\bm{\delta }} = \left[ {{\delta _0},{\delta _1},...,{\delta _{{L_\tau }{K_\nu } - 1}}} \right]^T \in {\mathbb{R}^{{L_\tau }{K_\nu }\times 1}}$ contains the hyperparameters that regulate the row sparsity of $\mathbf{\bar H}$ and is assumed to follow a Gamma distribution with the expression of 
\begin{equation}
    p\left( {\bm{\delta }} \right) = \prod\limits_{j = 0}^{{L_\tau }{K_\nu }-1} {\text{Gamma} \left( {{\delta _j};1,b} \right)},
\end{equation}
where the parameters $b$ are set empirically to encourage the sparsity of $\mathbf{\bar H}_{n_r}$. The Gamma distribution is defined to have the expression of $\text{Gamma} \left( {\delta _j;a,b} \right) = \Gamma {\left( a \right)^{ - 1}}{b^a}{\delta _j^{a - 1}}{\operatorname{e} ^{ - b\delta _j}}$, where $\Gamma \left( a \right) = \int_0^\infty  {{t^{a - 1}}{\operatorname{e} ^{ - t}}} dt$.

{In addition, the equivalent noise matrix $\mathbf{\bar W}^{\left(t\right)}$ contains both the additive Gaussian noise and the residual interference caused by data
detection errors. Following the commonly adopted equivalent Gaussian approximation in iterative inference frameworks \cite{GAN_10543050,SP_SBL_SIC_10640141}, $\mathbf{\bar W}^{\left(t\right)}$
is approximated as a complex Gaussian random matrix with the distribution of
\begin{equation}
    p\left( {{\mathbf{\bar W}}^{(t)}|\beta} \right) =  \prod_{n_r=0}^{N_r-1}\mathcal{C}\mathcal{N}\left( {{\mathbf{\bar W}^{(t)}_{n_r}};{\mathbf{0}},{{\beta} ^{ - 1}}{{\mathbf{I}}_{N}}} \right),
\end{equation}
where ${\mathbf{\bar W}^{(t)}_{n_r}}$ represents the $n_r$-th column of $\mathbf{\bar W}^{\left(t\right)}$ and $\beta$ is the inverse of the equivalent noise variance. }The hyperparameter $\beta$ is further assumed to follow a Gamma prior, expressed as
\begin{equation}
    p\left( \beta  \right) =\text{Gamma} \left( {\beta ;d,e} \right),
\end{equation}
where $d,e>0$ are small fixed constants. 

In practice, the resolutions of the virtual grid are finite. As a result, the true delay and Doppler shift values rarely coincide exactly with the predefined virtual grid points. We denote $\bm{\iota}$ and $\bm{\kappa}$ as the off-grid components corresponding to the delay grids $\bm{\bar \ell}$ and Doppler grids $\bm{\bar k}$, respectively, and these components share the same sparse support as $\mathbf{\bar H}$. With their boundedness known as a prior, elements in $\bm{\iota}$ and $\bm{\kappa}$ are assumed to follow a uniform prior distribution with the expression of 
\begin{equation}
    p\left( {{\iota _j}} \right) = U\left[ { - \frac{{{r_\tau}}}{2},\frac{{{r_\tau}}}{2}} \right],
\end{equation}
\begin{equation}
    p\left( {{\kappa _j}} \right) = U\left[ { - \frac{{{r_k}}}{2},\frac{{{r_k}}}{2}} \right],
\end{equation}

where $j \in {\left\{0,1,...,{{L_\tau }{K_\nu }-1}\right\}}$.

\section{Proposed GESBL for Data-aided ISAC}
To address the MMV sparse recovery problem in the presence of off-grid components, we propose an iterative channel estimation and sensing method that combines off-grid component estimation with a grid evolution refinement strategy, referred to as D-GESBL. 
\subsection{GESBL Framework}
{Different from the off-grid SBL scheme \cite{OGSBL9738478} with fixed virtual grids, we iteratively adjust the virtual grids by estimating the off-grid components based on the first-order Taylor approximation, so that the virtual grids are iteratively and non-uniformly refined according to the on-grid components and the off-grid components based on the current grids. Through the dynamic grid evolution strategy, the virtual grids can progressively match the real parameter values, which effectively mitigates the off-grid mismatch as well as the linearization error introduced by the first-order Taylor approximation, thereby enhancing the estimation performance. Each iteration consists of three step. In the first step, given the hyperparameters from the last iteration, the conditional posterior distribution is estimated. In the second step, with the estimated conditional posterior distribution, hyperparameters $\bm{\delta}$, ${\beta}$, $\bm{\kappa}$, and $\bm{\iota}$ are updated within the SBL framework by using expectation maximization (EM) algorithm. In the third step, the virtual grids are updated according to the second stage estimation.}

\textbf{Step One: Conditional Posterior Distribution Estimation.}
 Specifically, in the $g$-th iteration of channel estimation, given the virtual grids $\bm{\bar k}^{(g-1)}$ and $\bm{\bar \ell}^{(g-1)}$, we first formulate $\hat{\mathbf{\Phi }}^{\left(t\right)}\left( \bm{\bar k }^{\left(g-1\right)},\bm{\bar \ell }^{\left(g-1\right)}\right)$ according to \eqref{PHI_t} {with off-grid components $\bm{\kappa }^{(g-1)}$ and $\bm{\iota }^{(g-1)}$ set to be zero}. Subsequently, we denote the hyperparameter set by $\bm{\zeta }^{(g-1)}=\{\bm{\delta}^{(g-1)}, \beta^{(g-1)}\}$ for simplicity and given $\bm{\zeta }^{(g-1)}$, the conditional posterior distribution of $\mathbf{\bar H}$ can be expressed as
\begin{equation}\label{condi_post}
    \begin{split}
        &p\left( {\mathbf{\bar H}}|{\mathbf{Y}};\bm{\zeta }^{(g-1)}, \bm{\bar k }^{\left( g-1 \right)},\bm{\bar \ell}^{\left( g-1 \right)}\right)\\ 
        &\propto p\left( {\mathbf{Y}}|{\mathbf{\bar H}};\bm{\zeta }^{(g-1)},\bm{\bar k }^{\left( g-1 \right)},\bm{\bar \ell }^{\left( g-1 \right)} \right)p\left( {{\mathbf{\bar H}}|{{\bm{\delta }}^{\left( g-1 \right)}}} \right),
    \end{split}
\end{equation}
where the prior distribution $p\left( {{\mathbf{\bar H}}|{{\bm{\delta }}^{\left( g-1 \right)}}} \right)$ is given in \eqref{h_prior} and the likelihood distribution in the $g$-th iteration is given by
\begin{equation}\label{likely}
\begin{split}
    &p\left( {\mathbf{Y}}|{\mathbf{\bar H}};\bm{\zeta }^{(g-1)},\bm{\bar k }^{\left( g-1 \right)},\bm{\bar \ell }^{\left( g-1 \right)} \right) \\
    &= \prod_{n_r=0}^{N_r-1}\mathcal{C}\mathcal{N}\left( {\mathbf{Y}}_{n_r}; \hat{\mathbf{\Phi }}^{\left(t\right)}\left( \bm{\bar k }^{\left(g-1\right)},\bm{\bar \ell }^{\left(g-1\right)}\right){\mathbf{\bar H}_{n_r}},\frac{1}{{{\beta ^{\left( g -1\right)}}}}{{\mathbf{I}}} \right),
\end{split}
\end{equation}

By combining equations \eqref{h_prior}, \eqref{condi_post}, and \eqref{likely}, the conditional posterior distribution can be derived as

\begin{equation}
\begin{split}
    &p\left( {\mathbf{\bar H}}|{\mathbf{Y}};\bm{\zeta }^{(g-1)}, \bm{\bar k }^{\left( g-1 \right)},\bm{\bar \ell }^{\left( g-1 \right)}\right)\\
    &= \prod_{n_r=0}^{N_r-1}\mathcal{C}\mathcal{N}\left( {{\mathbf{\bar H}}_{n_r};{{\bm{\mu }}_{n_r}^{\left( g \right)}},{{\bm{\Sigma }}^{\left( g \right)}}} \right),
\end{split}
\end{equation}
with the posterior covariance matrix and the mean matrix of
\begin{equation}\label{sigma}
   {{\mathbf{\Sigma }}^{\left( g \right)}} = {\left( {{\beta ^{\left( g-1 \right)}}\hat{\mathbf{\Phi }}^{\left(t,g-1\right)H}
   \hat{\mathbf{\Phi }}^{\left(t,g-1\right)} + \left({\mathbf{\Delta }}^{(g-1)}\right)^{-1}} \right)^{ - 1}},
\end{equation}
and
\begin{equation}\label{miu}
    {{\bm{\mu }}^{\left( g \right)}} = {\beta ^{\left( g-1 \right)}}{{\mathbf{\Sigma }}^{\left( g \right)}}\hat{\mathbf{\Phi }}^{\left(t,g-1\right)H}{\mathbf{Y}},
\end{equation}
where $\hat{\mathbf{\Phi }}^{\left(t,g-1\right)}$ is the simplified expression of $\hat{\mathbf{\Phi }}^{\left(t\right)}\left( \bm{\bar k}^{\left(g-1\right)},\bm{\bar \ell }^{\left(g-1\right)}\right)$ and ${\bm{\mu }}^{\left( g \right)}\in \mathbb{C}^{N\times N_r}=\left[{\bm{\mu }}^{\left( g \right)}_{0},{\bm{\mu }}^{\left( g \right)}_{1},...,{\bm{\mu }}^{\left( g \right)}_{N_r-1}\right]$. 

\textbf{Step Two: Hyperparameters Update.}
Then, given the conditional posterior distribution, the hyperparameters $\bm{\delta}^{(g)}$, ${\beta}^{(g)}$ can be updated by maximizing the log joint distribution ${\log \left( {p\left( {{\mathbf{\bar H}},{\bm{\delta }},\beta, {\mathbf{Y}}} \right)} \right)}$ via the EM algorithm. In the expectation step, we evaluate $Q\left(\bm{\zeta}|{{\bm{\zeta }}^{\left( g-1 \right)}},\bm{\bar k }^{\left( g-1 \right)},\bm{\bar \ell }^{\left( g-1 \right)} \right)$ as the cost function, which is obtained by taking the expectation of the log joint distribution ${\log \left( {p\left( {{\mathbf{\bar H}},{\bm{\zeta }},{\mathbf{Y}}} \right)} \right)}$ under the conditional posterior $p\left( {\mathbf{\bar H}}|{\mathbf{Y}};\bm{\zeta}^{(g-1)},\bm{\bar k}^{\left( g-1 \right)},\bm{\bar \ell}^{\left( g-1 \right)} \right)$ with an expression of
\begin{equation}
    \begin{split}
        &Q\left( {\bm{\zeta }}|{{\bm{\zeta }}^{\left( g-1 \right)}},\bm{\bar k }^{\left( g-1 \right)},\bm{\bar \ell }^{\left( g-1 \right)} \right) \\
        &= {\mathbb{E}_{{\mathbf{\bar H}}|{\mathbf{Y}};{{\bm{\zeta }}^{\left( g-1 \right)}},\bm{\bar k }^{\left( g-1 \right)},\bm{\bar \ell}^{\left( g-1 \right)}}}\left\{ {\log \left( {p\left( {{\mathbf{\bar H}},{\bm{\zeta }},{\mathbf{Y}}} \right)} \right)} \right\}.
    \end{split}
    \label{Q}
\end{equation}
In the maximization step, the hyperparameters $\bm{\delta}$ and $\beta$ are updated through maximizing \eqref{Q}, i.e.,
\begin{equation}\label{maxQ_delta}
    {{\bm{\delta }}^{\left( {g} \right)}} = \mathop {\arg \max }\limits_{\bm{\delta }} Q\left( {\bm{\delta }}|{{\bm{\zeta }}^{\left( g-1 \right)}},\bm{\bar k }^{\left( g-1 \right)},\bm{\bar \ell }^{\left( g-1 \right)} \right),
\end{equation}
\begin{equation}\label{maxQ_beta}
    {\beta ^{\left( {g } \right)}} = \mathop {\arg \max }\limits_{\beta} Q\left( {\beta}|{{\bm{\zeta }}^{\left( g-1 \right)}},\bm{\bar k }^{\left( g-1 \right)},\bm{\bar \ell }^{\left( g-1 \right)} \right).
\end{equation}
\begin{figure*}
\begin{equation}
\label{delta}
\delta _j^{\left( {g} \right)} = \frac{{\sqrt {N_r^2 + 4b\sum_{n_r=0}^{N_r-1}\left( {{{\left| {{\mathbf{\mu }}_{j,n_r}^{\left( g \right)}} \right|}^2} + {\mathbf{\Sigma }}_{j,j}^{\left( g \right)}} \right)}  - N_r}}{{2b}}
\end{equation}
\begin{equation}
\label{beta}
    {\beta ^{\left( {g} \right)}} 
    = \frac{{d - 1 + N_r{L_\tau }{K_\nu }}}{{e + {{\left\| {{\mathbf{Y}} - \hat{\mathbf{\Phi }}^{\left(t\right)}\left( \bm{\bar k }^{\left(g-1\right)},\bm{\bar \ell }^{\left(g-1\right)} \right){{\bm{\mu }}^{\left( g \right)}}} \right\|}_F^2} + \frac{N_r}{{{\beta ^{\left( g-1 \right)}}}}\sum\limits_{j = 0}^{{L_\tau }{K_\nu }-1} {\left( {1 - \frac{{{\mathbf{\Sigma }}_{j,j}^{\left( g \right)}}}{{\delta _j^{\left( g-1 \right)}}}} \right)} }}
\end{equation}
\vspace{-0.15in}
\end{figure*}

{Proposition 1: 
The closed-form updating rule for $\bm{\delta}$ and $\beta$ can be given by \eqref{delta} and \eqref{beta} {on the top of the next page}, respectively.}

\textit{{Proof: Please refer to Appendix A.}}

For updating the off-grid components, we apply the first-order Taylor expansion to the measurement matrix, yielding
\begin{equation}
\begin{split}
    \tilde{{\mathbf{\Phi }}}^{\left(t\right)}\left( \bm{\bar k },\bm{\bar \ell },\bm{\kappa},{\bm{\iota }} \right) =& {\mathbf{\hat{\mathbf{\Phi }}}^{\left(t\right)}\left( \bm{\bar k },\bm{\bar \ell } \right)} + {\mathbf{\hat{\mathbf{\Phi }}}_\mathbf{B}^{\left(t\right)}\left( \bm{\bar k},\bm{\bar \ell } \right)}{\text{diag}}\left( \bm{\kappa}  \right)\\
    &+ {\mathbf{\hat{\mathbf{\Phi }}}_\mathbf{C}^{\left(t\right)}\left( \bm{\bar k },\bm{\bar \ell } \right)}{\text{diag}}\left( \bm{\iota}  \right),\label{PHI_t_taylor}
\end{split} 
\end{equation}
where $\mathbf{\hat{\mathbf{\Phi }}_B}^{\left(t\right)} = \left[ {{\mathbf{\hat b}}^{\left(t\right)}\left( {{{\bar \ell}_0},{{\bar k}_0}} \right),...,{\mathbf{\hat b}}^{\left(t\right)}\left( {{{\bar \ell}_{{L_\tau }{K_\nu } - 1}},{{\bar k}_{{L_\tau }{K_\nu } - 1}}} \right)} \right]$, $\mathbf{\hat{\mathbf{\Phi }}_C}^{\left(t\right)} = \left[ {{\mathbf{\hat c}}^{\left(t\right)}\left( {{{\bar \ell}_0},{{\bar k}_0}} \right),...,{\mathbf{\hat c}}^{\left(t\right)}\left( {{{\bar \ell}_{{L_\tau }{K_\nu } - 1}},{{\bar k}_{{L_\tau }{K_\nu } - 1}}} \right)} \right]$, $\bm{\iota}  = {\left[ {{\iota _0},{\iota _1},...,{\iota _{{L_\tau }{K_\nu } - 1}}} \right]^T}$, and $\bm{\kappa}  = {\left[ {{\kappa _0},{\kappa _1},...,{\kappa _{{L_\tau }{K_\nu } - 1}}} \right]^T}$. 
${\mathbf{\hat b}}^{\left(t\right)}\left( {{{\bar \ell}_{{j}}},{{\bar k}_{{j}}}} \right) = \frac{{{\partial\mathbf{\hat a}}^{\left(t\right)}\left( {{{\bar \ell}_{{j}}},{{\bar k}_{{j}}}} \right)}}{{\partial {{\bar \ell}_{{j}}}}}$ and ${\mathbf{\hat c}}^{\left(t\right)}\left( {{{\bar \ell}_{{j}}},{{\bar k}_{{j}}}} \right) = \frac{{{\partial\mathbf{\hat a}}^{\left(t\right)}\left( {{{\bar \ell}_{{j}}},{{\bar k}_{{j}}}} \right)}}{{\partial {{\bar k}_{{j}}}}}$ with ${\mathbf{\hat a}}^{\left(t\right)}\left( {{{\bar \ell}_{{j}}},{{\bar k}_{{j}}}} \right)=\mathbf{\Xi }\left( {{{\bar \ell}_{{j}}},{{\bar k}_{{j}}}} \right)\bm{\hat x}^{\left(t\right)}$ being the $j$-th column of ${\mathbf{\hat{\mathbf{\Phi }}}^{\left(t\right)}\left( \bm{\bar k },\bm{\bar \ell } \right)}$.

Proposition 2: Based on \eqref{PHI_t_taylor}, the off-grid components $\bm{\kappa}$ and $\bm{\iota}$ can be jointly updated by \eqref{E_kappa}, which is shown on the top of the next page.


\begin{figure*}
    \begin{equation}\label{E_kappa}
    {\bm{\kappa} ^{\left( {g } \right)}},{\bm{\iota} ^{\left( {g } \right)}}= \mathop {\arg \min }\limits_{{\bm{\kappa}},{\bm{\iota}}} {\mathbb{E}_{{\mathbf{\bar H}}|{\mathbf{Y}};{{\bm{\zeta }}^{\left( g-1 \right)}},\bm{\bar k }^{\left( g-1 \right)},\bm{\bar \ell}^{\left( g-1 \right)}}}\left( \left\|\mathbf{Y}- \tilde{{\mathbf{\Phi }}}^{\left(t\right)}\left( \bm{\bar k }^{(g-1)},\bm{\bar \ell }^{(g-1)},\bm{\kappa},{\bm{\iota }} \right) \mathbf{\bar H}\right\|_F^2\right)
\end{equation}

\noindent\rule{\textwidth}{0.6pt}  
\end{figure*}
The $j$-th entry of the off-grid components $\bm{\iota}$ and $\bm{\kappa}$ can be updated by
\begin{equation}\label{kappa}
    {\kappa} _j^{\left( {g} \right)} = \frac{{{\alpha} _j^{\kappa\left( g \right)} - \left\{ {{\bm{Z} }_j^{\kappa\left( g \right)}} \right\}_{ - j}^T{{\left\{ {{{\bm{\kappa }}^{\left( g-1 \right)}}} \right\}}_{ - j}}}}{{{Z} _{j,j}^{\kappa\left( g \right)}}},
\end{equation}
\begin{equation}\label{iota}
    {\iota} _j^{\left( {g} \right)} = \frac{{{\alpha} _j^{\iota\left( g \right)} - \left\{ {{\bm{Z} }_j^{\iota\left( g \right)}} \right\}_{ - j}^T{{\left\{ {{{\bm{\iota }}^{\left( g-1 \right)}}} \right\}}_{ - j}}}}{{{Z} _{j,j}^{\iota\left( g \right)}}},
\end{equation}
where ${{\bm{Z} }_j^{\kappa\left( g \right)}}$ and ${{\bm{Z} }_j^{\iota\left( g \right)}}$ are the $j$-th column of matrices ${{\bm{Z} }^{\kappa\left( g \right)}}$ and ${{\bm{Z} }^{\iota\left( g \right)}}$, respectively. $ \left\{  \cdot  \right\}_{ - j}$ denotes the vector without the $j$-th entry. ${{\bm{Z} }^{\kappa\left( g \right)}}$ and ${{\bm{Z} }^{\iota\left( g \right)}}$ have the expression of 
\begin{equation}
    {{\mathbf{Z}}^{\kappa\left( g \right)}} = \operatorname{Re} \left\{ {{{\mathbf{\hat{\mathbf{\Phi }}}_\mathbf{B}^{\left(t,g-1\right)H}}}{\mathbf{\hat{\mathbf{\Phi }}}_\mathbf{B}^{\left(t,g-1\right)}} \circ \left( {N_r{\mathbf{\Sigma }}^{\left( g \right)} + {{\bm{\mu }}^{\left( g \right)}}{{\bm{\mu }}^{\left( g \right)H}}} \right)} \right\},
\end{equation}
\begin{equation}
    {{\mathbf{Z}}^{\iota\left( g \right)}} = \operatorname{Re} \left\{ {{{\mathbf{\hat{\mathbf{\Phi }}}_\mathbf{C}^{\left(t,g-1\right)H}}}{\mathbf{\hat{\mathbf{\Phi }}}_\mathbf{C}^{\left(t,g-1\right)}} \circ \left( {N_r{\mathbf{\Sigma }}^{\left( g \right)} + {{\bm{\mu }}^{\left( g \right)}}{{\bm{\mu }}^{\left( g \right)H}}} \right)} \right\},
\end{equation}
and the expressions of ${{\bm{\alpha} }^{\kappa\left( g \right)}}$ and ${{\bm{\alpha} }^{\iota\left( g \right)}}$ are shown in \eqref{alpha_kappa} and \eqref{alpha_iota} on the bottom of the next page,
where ${\bf{\hat \Phi }}^{\left( {t,g - 1} \right)}$, ${\bf{\hat \Phi }}_{\bf{B}}^{\left( {t,g - 1} \right)}$, and ${\bf{\hat \Phi }}_{\bf{C}}^{\left( {t,g - 1} \right)}$ are the simplified expression of ${\bf{\hat \Phi }}^{\left( {t} \right)}\left(\bm{\bar k}^{(g-1)},\bm{\bar \ell}^{(g-1)}\right)$, ${\bf{\hat \Phi }}_{\bf{B}}^{\left( {t} \right)}\left(\bm{\bar k}^{(g-1)},\bm{\bar \ell}^{(g-1)}\right)$, and ${\bf{\hat \Phi }}_{\bf{C}}^{\left( {t} \right)}\left(\bm{\bar k}^{(g-1)},\bm{\bar \ell}^{(g-1)}\right)$, respectively.\\
\textit{Proof: Please refer to Appendix B.}
\begin{figure*}[!b]
\noindent\rule{\textwidth}{0.6pt}  
\begin{equation}\label{alpha_kappa}
\begin{split}
    {{\bm{\alpha }}^{\kappa\left( g \right)}} = &
    \operatorname{Re} \left\{\sum\limits_{n_r = 0}^{N_r-1}{\text{diag}}\left( {{{\bm{\mu }}_{n_r}^{\left( g \right)H}}} \right){{\mathbf{\hat{\mathbf{\Phi }}}_\mathbf{B}^{\left(t,g-1\right)H}}}\left( {{\mathbf{Y}} - {\mathbf{\hat{\mathbf{\Phi }}}^{\left(t,g-1\right)}}{\bm{\mu }_{n_r}^{\left( g \right)}}} \right) \right\}-N_r\operatorname{Re}\left\{{\text{diag}}\left( {{{\mathbf{\hat{\mathbf{\Phi }}}_\mathbf{B}^{\left(t,g-1\right)H}}}{\mathbf{\hat{\mathbf{\Phi }}}^{\left(t,g-1\right)}}{{\mathbf{\Sigma }}^{\left( g \right)}}} \right) \right\}.
\end{split}
\end{equation}
\textcolor{black}{
\begin{equation}\label{alpha_iota}
\begin{split}
    {{\bm{\alpha }}^{\iota\left( g \right)}}=&\operatorname{Re} \left\{ \sum\limits_{n_r = 0}^{N_r-1}{\text{diag}}\left( {{{\bm{\mu }}_{n_r}^{\left( g \right)H}}} \right){{\mathbf{\hat{\mathbf{\Phi }}}_\mathbf{C}^{\left(t,g-1\right)H}}}\left({{\mathbf{Y}} - {\mathbf{\hat{\mathbf{\Phi }}}^{\left(t,g-1\right)}}{\bm{\mu }_{n_r}^{\left( g \right)}}} \right) \right\}- N_r\operatorname{Re}\left\{{\text{diag}}\left( {{{\mathbf{\hat{\mathbf{\Phi }}}_\mathbf{C}^{\left(t,g-1\right)H}}}{\mathbf{\hat{\mathbf{\Phi }}}^{\left(t,g-1\right)}}{{\mathbf{\Sigma }}^{\left( g \right)}}} \right) \right\}\\
    &-\operatorname{Re} \left\{ {{{\mathbf{\hat{\mathbf{\Phi }}}_\mathbf{B}^{\left(t,g-1\right)H}}}{\mathbf{\hat{\mathbf{\Phi }}}_\mathbf{C}^{\left(t,g-1\right)}} \circ \left( {N_r{\mathbf{\Sigma }}^{\left( g \right)} + {{\bm{\mu }}^{\left( g \right)}}{{\bm{\mu }}^{\left( g \right)H}}} \right)} \right\}\bm{\kappa}^{(g)}.
\end{split}
\end{equation}}
\end{figure*}

Due to the fact that $\bm{\iota}$ and $\bm{\kappa}$ share the same sparse support with $\bm{\delta}$, limited entries of $\bm{\iota}$ and $\bm{\kappa}$ need to be updated for complexity reduction. {Although the number of non-zero rows of $\mathbf{\bar H}$ is unknown, there are at most $\bar R=\left\lfloor {\frac{N}{{\log \left( {{L_\tau }{K_\nu }} \right)}}} \right\rfloor$ non-zero entries can be recovered in this system \cite{OGSBL9738478}. Therefore, we assume there are $\bar R$ non-zero entries for $\bm{\iota}$ and $\bm{\kappa }$ when updating them, and their corresponding index set is given by $\mathcal{R}$. Then, $\bm{\kappa }$, $\bm{\iota }$, $\bm{\alpha}^{\kappa}$, $\bm{\alpha}^{\iota}$ and ${\bm{Z} }^{\kappa}$ and ${\bm{Z} }^{\iota}$ are truncated to $\bar{\bm{\kappa }} \in {\mathbb{R}^{\bar R\times 1}}$, $\bar{\bm{\iota }} \in {\mathbb{R}^{\bar R\times 1}}$,  $\bm{\bar\alpha}^{\kappa} \in {\mathbb{R}^{\bar R\times 1}}$, $\bm{\bar\alpha}^{\iota} \in {\mathbb{R}^{\bar R\times 1}}$, ${\bm{\bar Z} }^{\kappa} \in {\mathbb{R}^{\bar R\times \bar R}}$ and ${\bm{\bar Z} }^{\iota} \in {\mathbb{R}^{\bar R\times \bar R}}$, respectively.
The remaining entries of $\bm{\kappa}$ and $\bm{\iota}$ keep to be zero without updating.}

\textbf{Step Three: Update of the Virtual Grids.} 
Finally, the dynamic grids in the $g$-the iteration can be adjusted by
\begin{equation}\label{k_update}
    \bm{\bar k}^{(g)}=\bm{\bar k}^{(g-1)}+\bm{\kappa}^{(g)}
\end{equation}
and
\begin{equation}\label{l_update}
    \bm{\bar \ell}^{(g)}=\bm{\bar \ell}^{(g-1)}+\bm{\iota}^{(g)}.
\end{equation}

The iterative process of our proposed GESBL for data-aided channel estimation and target sensing scheme for the AFDM-ISAC system is summarized in \textbf{Algorithm \ref{GE-SBL}}. 

\subsection{Channel and Sensing-related Parameter Estimation}
After obtaining all the channel-related parameters, the effective channel response in the DAF domain for the $n_r$-the receive antenna can be expressed as
\begin{equation}\label{H_eff_est}
    \mathbf{\hat H}_{eff,n_r}^{(t)}= \sum\limits_{r \in \mathcal{R}} {\mu}_{r,n_r}\mathbf{A}{{\mathbf{\Delta }}_{{\bar k_r}}}{{\mathbf{\Pi }}^{\bar \ell_r}}\mathbf{A}^H,
\end{equation}
with $n_r = 0,...,N_r-1$. 
{All the estimated channel-related parameters can be considered as prior information in the next data-aided iteration to accelerate convergence.}

{For the sensing-related parameters, it is necessary to determine the number of targets, as well as their AoAs, delays for target localization, and Doppler shifts for target velocity estimation. Specifically,
we extract the index $\hat r \in \mathcal{\hat R}$ with size $\hat R$, which satisfies $\sum\limits_{n_r=0}^{N_r-1}{\mu}_{\hat r,n_r}>\varepsilon_{thr}$ to determine the number of targets, where ${\mu}_{\hat r,n_r}$ is the $\left({\hat r,n_r}\right)$-th element of $\bm{\mu}$.} Then, the delay and Doppler shift are estimated as $\hat{\bm{\eta}}=\bm{\bar \ell}_{\mathcal{\hat R}}$ and $\hat{\bm{\nu}}=\bm{\bar k}_{\mathcal{\hat R}}$, respectively. Also, for the AoA estimation, we define the submarix $\bm{\hat \mu}_{\mathcal{\hat R}}\in \mathbb{C}^{\hat R \times N_r}$, with the expression of
\begin{equation}
    \bm{\hat \mu}_{\mathcal{\hat R}} = \bm{\mu}\left(\mathcal{\hat R},:\right).
\end{equation}
Thus, the AoA of target $i$ can be estimated easily by a one dimensional process, with the expression of
\begin{equation}\label{AoA_est}
    {\hat \theta }_i = \arg \mathop {\max }\limits_{\theta}\frac{{\left| {{\bm{\hat \mu}_{\mathcal{\hat R}}}\left(i,:\right){{\bf{a}}^H_R}\left( \theta  \right)} \right|}}{{{{\left\| {\bm{\hat \mu}_{\mathcal{\hat R}}}\left(i,:\right) \right\|}_2}{{\left\| {{{\bf{a}}_R}\left( \theta  \right)} \right\|}_2}}},
\end{equation}
\vspace{-0.1in}
where $i=1,...,\hat R$.

\subsection{Complexity Analysis}
The complexity of the proposed GESBL for data-aided framework in Algorithm \ref{GE-SBL} is analyzed as follows. In each iteration, the update of $\hat{\mathbf{\Phi }}\left( \bm{\bar \ell },\bm{\bar k } \right)$ has the complexity of $\mathcal{O}\left(\bar RNL_\tau K_\nu\right)$, and the estimation of $\bm{\mu}$ and $\bm{\Sigma}$ have the complexity of $\mathcal{O}\left(NL^2_\tau K^2_\nu +L_\tau K_\nu N N_r\right)$ and $\mathcal{O}\left(NL^2_\tau K^2_\nu+L_\tau K_\nu+L^3_\tau K^3_\nu\right)$, respectively. The update of the hyperparameters $\bm{\delta}$ and $\beta$ have the complexity of $\mathcal{O}\left(L_\tau K_\nu\right)$ and $\mathcal{O}\left(NL_\tau K_\nu N_r+NN_r+L_\tau K_\nu\right)$, respectively. The updates of the off-grid components $\bm{\kappa}$ and $\bm{\iota}$ both have the complexity of $\mathcal{O}\left(\bar R^3+\bar R^2\right)$. 
It can be observed that the dominant computational complexity lies in the matrix inversion required for variance computation, leading to a relatively high computational burden. Therefore, an inverse-free low complexity version is required, which is introduced in the subsequent section.
\begin{algorithm}
\caption{Proposed GESBL for channel parameter sensing with $t$-th data-aided iteration}\label{GE-SBL}
\KwIn{${\mathbf{Y}}$, parameters $b$, $d$, $e$, the convergence tolerance $\varepsilon_g$, maximum iteration $I_{\max }$}
\textbf{Initialization}:
$g=1$, $\boldsymbol{\bar k}^{(0)}$, $\bar{\boldsymbol{\ell}}^{(0)}$,${\beta ^{\left( 0 \right)}}$, ${\bm{\delta }}^{\left( 0 \right)}$, ${\bm{\kappa }}^{\left( 0 \right)} = {\mathbf{0}}$, ${\bm{\iota}}^{\left( 0 \right)} = {\mathbf{0}}$\;
\textbf{Repeat}

Formulate $\hat{\mathbf{\Phi }}^{\left(t\right)}\left( \bm{\bar k }^{\left(g-1\right)},\bm{\bar \ell }^{\left(g-1\right)} \right)$ according to \eqref{PHI_t}\;
Compute ${{\mathbf{\Sigma }}^{\left( g \right)}}$ and ${{\bm{\mu }}^{\left( g \right)}}$ according to \eqref{sigma} and \eqref{miu}, respectively\;
Update the hyper-parameters ${{\bm{\delta }}^{\left( {g} \right)}}$, ${\beta ^{\left( {g } \right)}}$, ${{\bm{\kappa }}^{\left( {g} \right)}}$, and ${{\bm{\iota }}^{\left( {g} \right)}}$ according to \eqref{delta}, \eqref{beta}, \eqref{kappa} and \eqref{iota}, respectively\;
Update ${\bar{\bm{k }}^{\left( {g} \right)}}$ and ${\bar{\bm{\ell}}^{\left( {g } \right)}}$ according to \eqref{k_update} and \eqref{l_update}, and set the off gird components $\bm{\kappa}^{(g)}=\bm{\iota}^{(g)}=\mathbf{0}$\;
$g\leftarrow g+1$\;
\textbf{until} $ \frac{{\left\| {{{\bm{\delta }}^{\left( {g } \right)}} - {{\bm{\delta }}^{\left( g-1 \right)}}} \right\|^2}}{{\left\| {{{\bm{\delta }}^{\left( g-1 \right)}}} \right\|^2}} < \varepsilon_g $ or $g=I_{\max }$\;
$\bm{\mu}=\bm{\mu}^{(g)}$, $\bm{\bar k}=\bm{\bar k}^{(g)}$, $\bm{\bar \ell}=\bm{\bar \ell}^{(g)}$, $\bm{\delta}=\bm{\delta}^{(g)}$, $\beta^{(g)}=\beta$\;
\KwOut{$\bm{\mu}$, $\bm{\bar k}$, $\bm{\bar \ell}$, $\bm{ \delta}$, $\beta$}
\end{algorithm}

\begin{algorithm}
\caption{Proposed GAMP-GESBL for channel parameter sensing with $t$-th data-aided iteration}\label{GAMP-GE-SBL}
\KwIn{${\mathbf{Y}}$, parameters $b$, $d$, $e$, $\varepsilon_g$, $\varepsilon_q$, $I_{\max }$, $Q_{\max }$}
\textbf{Initialization}:
$g=q=1$, $\bm{\bar k}^{(0)}$, $\bar{\bm{\ell}}^{(0)}$,${\beta ^{\left( 0 \right)}}$, ${\bm{\delta }}^{\left( 0 \right)}$, $\bm{\mu}^{\bar H (0)}$, $\mathbf{V}^{\bar H (0)}$, $\forall m, n_r$, $S_{m,n_r}^{(0)}=0$, ${\bm{\kappa }}^{\left( 0 \right)} = {\mathbf{0}}$, ${\bm{\iota }}^{\left( 0 \right)} = {\mathbf{0}}$\;
\textbf{Repeat}

Formulate $\hat{\mathbf{\Phi }}^{\left(t\right)}\left( \bm{\bar k }^{\left(g-1\right)},\bm{\bar \ell }^{\left(g-1\right)} \right)$ according to \eqref{PHI_t}\;
\textbf{For} $q<Q_{max}$ and $\frac{{\sum\limits_{j = 0}^{{L_\tau }{N_\nu } - 1} {\sum\limits_{{n_r} = 0}^{{N_r-1}} {{{\left| {\mu _{j,{n_r}}^{\bar H\left( q \right)} - \mu _{j,{n_r}}^{\bar H\left( {q - 1} \right)}} \right|}^2}} } }}{{\sum\limits_{j = 0}^{{L_\tau }{N_\nu } - 1} {\sum\limits_{{n_r} = 0}^{{N_r-1}} {{{\left| {\mu _{j,{n_r}}^{\bar H\left( {q - 1} \right)}} \right|}^2}} } }}<\varepsilon_q$\\
$\forall m, n_r,$ $Z_{m,{n_r}}^{\left( q \right)} = \sum\limits_{j = 0}^{{L_\tau }{K_\nu }-1} {\hat \Phi _{m,j}^{\left( {t,g-1} \right)}} \mu _{j,{n_r}}^{\bar H\left( {q - 1} \right)}$\;
$V_{m,{n_r}}^{P\left( q \right)} = {\sum\limits_{j = 0}^{{L_\tau }{K_\nu }-1} {\left| {\hat \Phi _{m,j}^{\left( {t,g-1} \right)}} \right|} ^2}V_{j,{n_r}}^{\bar H\left( {q - 1} \right)}$\;
$\forall m, n_r,$ $P_{m,{n_r}}^{\left( q \right)} = Z_{m,{n_r}}^{\left( q \right)} - V_{m,{n_r}}^{P\left( q \right)}S_{m,{n_r}}^{\left( {q - 1} \right)}$\;
$\forall m, n_r,$ $S_{m,{n_r}}^{\left( q \right)} = \left( {1 - {\vartheta _s}} \right)S_{m,{n_r}}^{\left( {q - 1} \right)} + {\vartheta _s}{g_S}\left( {P_{m,{n_r}}^{\left( q \right)},V_{m,{n_r}}^{P\left( q \right)},{\bm{\zeta}^{\left( g-1 \right)}}} \right)$\;
$\forall m, n_r,$ $V_{m,{n_r}}^{S\left( q \right)} = V_{m,{n_r}}^{P\left( q \right)}\frac{{\partial {g_S}\left( {P_{m,{n_r}}^{\left( q \right)},V_{m,{n_r}}^{P\left( q \right)},{\bm{\zeta}^{\left( g-1 \right)}}} \right)}}{{\partial {P_{m,{n_r}}}}}$\;
$\forall j, n_r,$ $V_{j,{n_r}}^{U\left( q \right)} = {\left( {{{\sum\limits_{m = 0}^{N-1 }{\left| {\hat \Phi _{m,j}^{\left( {t,g-1} \right)}} \right|} }^2}V_{m,{n_r}}^{S\left( q \right)}} \right)^{ - 1}}$\;
$\forall j, n_r,$ $U_{j,{n_r}}^{\left( q \right)} = \mu _{j,{n_r}}^{\bar H\left( {q - 1} \right)} + V_{j,{n_r}}^{U\left( q \right)}\sum\limits_{m = 1}^N {\hat \Phi _{m,j}^{\left( {t,g-1} \right)}} S_{m,{n_r}}^{\left( q \right)}$\;
$\forall j, n_r,$ $\mu _{j,{n_r}}^{\bar H\left( q \right)} = \left( {1 - {\vartheta _h}} \right)\mu _{j,{n_r}}^{\bar H\left( {q - 1} \right)} + {\vartheta _h}{g_{\bar H}}\left( {U_{j,{n_r}}^{\left( q \right)},V_{j,{n_r}}^{U\left( q \right)},{\bm{\zeta}^{\left( g-1 \right)}}} \right)$\;
$\forall j, n_r,$ $V_{j,{n_r}}^{\bar H\left( q \right)} = V_{j,{n_r}}^{U\left( q \right)}\frac{{\partial {g_{\bar H}}\left( {U_{j,{n_r}}^{\left( q \right)},V_{j,{n_r}}^{U\left( q \right)},{\bm{\zeta}^{\left( g-1 \right)}}} \right)}}{{\partial {U_{j,{n_r}}}}}$\;
\textbf{end}\;
$\bm{\mu}^{(g)}=\bm{\mu}^{\bar H (q)}$, ${{\mathbf{\Sigma }}^{\left( g \right)}}=\text{diag}\left(\frac{1}{N_r}{\sum \limits_{n_r=1}^{N_r}{\mathbf{V}_{n_r}^{\bar H(q)}}}\right)$\;
Update the hyper-parameters ${{\bm{\delta }}^{\left( {g} \right)}}$, ${\beta ^{\left( {g } \right)}}$, ${{\bm{\iota }}^{\left( {g} \right)}}$ and ${{\bm{\kappa }}^{\left( {g} \right)}}$ according to \eqref{delta}, \eqref{beta}, \eqref{iota} and \eqref{kappa}, respectively\;
Update ${\bar{\bm{\ell }}^{\left( {g} \right)}}$ and ${\bar{\bm{k }}^{\left( {g } \right)}}$ according to \eqref{l_update} and \eqref{k_update}, and set the off gird components $\bm{\kappa}^{(g)}=\bm{\iota}^{(g)}=\mathbf{0}$\;
$g\leftarrow g+1$\;
\textbf{until} $ \frac{{\left\| {{{\bm{\delta }}^{\left( {g} \right)}} - {{\bm{\delta }}^{\left( g-1 \right)}}} \right\|_F^2}}{{\left\| {{{\bm{\delta }}^{\left( g-1 \right)}}} \right\|_F^2}} < \varepsilon $ or $g=I_{\max }$\;
$\bm{\mu}=\bm{\mu}^{(g)}$, $\bm{\bar \ell}=\bm{\bar \ell}^{(g)}$, $\bm{\bar k}=\bm{\bar k}^{(g)}$, $\bm{\delta}=\bm{\delta}^{(g)}$, $\beta^{(g)}=\beta$\;
\KwOut{$\bm{\mu}$, $\bm{\bar k}$, $\bm{\bar \ell}$, $\bm{ \delta}$, $\beta$}
\end{algorithm}
\section{Proposed Low Complexity GAMP-GESBL for Data-aided ISAC}
In this section, we give a brief introduction of our proposed low complexity data-aided AFDM-ISAC scheme by using GAMP-GESBL framework. Developed in {\cite{GAMP6033942}}, the GAMP algorithm offers a computationally efficient way to approximate the messages in loopy belief propagation by employing Gaussian approximations derived from the central limit theorem and first-order Taylor expansion. These approximations transform the multi-dimensional message updates into separable scalar Gaussian operations, while still capturing the key posterior statistics required by Bayesian inference.
Motivated by this property, we develop the GAMP-based mean-and-variance estimation scheme instead of the conventional mean-and-variance computation of the conditional posterior distribution in \eqref{sigma} and \eqref{miu}.
In the $g$-th iteration of channel estimation stage, we also denote the hyperparameter set by $\bm{\zeta}^{(g-1)}=\{\bm{\delta}^{(g-1)}, \beta^{(g-1)}\}$ for simplicity. Treating these hyperparameters as known and fixed, and letting $\bar H_{j,n_r}$ denote the $(j,n_r)$-th entry of $\mathbf{\bar H}$, the posterior distribution can be approximated as
\begin{equation}
\begin{split}
    &p\left( {{{\bar H}_{j,{n_r}}}|{\bf{Y}},{U_{j,{n_r}}},V_{j,{n_r}}^U,{\bm{\zeta}^{\left( g-1 \right)}}} \right) \\
    &\propto p\left( {{{\bar H}_{j,{n_r}}}|{\bm{\zeta}^{\left( g-1 \right)}}} \right){\cal C}{\cal N}\left( {{{\bar H}_{j,{n_r}}};{U_{j,{n_r}}},V_{j,{n_r}}^U} \right)\\
    &={\cal C}{\cal N}\left( {{{\bar H}_{j,{n_r}}};\mu _{j,{n_r}}^{\bar H},V_{j,{n_r}}^{\bar H}} \right),
\end{split}
\end{equation}
with 
\begin{equation}
   \mu _{j,{n_r}}^{\bar H} = \frac{{{U_{j,{n_r}}}}}{{{\delta^{(g-1)} _j}V_{j,{n_r}}^U + 1}},
\end{equation}
\begin{equation}
V_{j,{n_r}}^{\bar H} = \frac{{V_{j,{n_r}}^U}}{{{\delta^{(g-1)} _j}V_{j,{n_r}}^U + 1}},
\end{equation}
where ${U_{j,{n_r}}}$ serves as a Gaussian-noise–corrupted approximation to ${\bar H_{j,{n_r}}}$ and ${V_{j,{n_r}}^U}$ is the effective noise variance. We write $\mathbf{Z} = \hat{\mathbf{\Phi }}^{\left(t,g-1\right)}\mathbf{\bar H}\in \mathbb{C}^{N \times N_r}$ as the noise-free output and denote $Z_{m,n_r}$ as the $(m,n_r)$-th entry of $\mathbf{Z}$. Then, the posterior distribution of the latent variable $Z_{m,n_r}$ can be approximated as
\begin{equation}
\begin{split}
   &p\left( {{Z_{m,{n_r}}}|{\bf{Y}},{P_{m,{n_r}}},V_{m,{n_r}}^P,{\bm{\zeta}^{\left( g-1 \right)}}} \right)\\
   &\propto p\left( {{Y_{m,{n_r}}}|{Z_{m,{n_r}}},{\bm{\zeta}^{\left( g-1 \right)}}} \right){\cal C}{\cal N}\left( {{Z_{m,{n_r}}};{P_{m,{n_r}}},V_{m,{n_r}}^P} \right)\\
   &={\cal C}{\cal N}\left( {{Z_{m,{n_r}}};\mu _{m,{n_r}}^Z,V_{m,{n_r}}^Z} \right), 
\end{split}
\end{equation}
with 
\begin{equation}
\mu _{m,{n_r}}^Z = \frac{{V_{m,{n_r}}^P{\beta ^{\left( g-1 \right)}}{Y_{m,{n_r}}} + {P_{m,{n_r}}}}}{{{\beta ^{\left( g \right)}}V_{m,{n_r}}^P + 1}},
\end{equation}
\begin{equation}
   V_{m,{n_r}}^Z = \frac{{V_{m,{n_r}}^P}}{{{\beta ^{\left( g-1 \right)}}V_{m,{n_r}}^P + 1}},
\end{equation}
where $P_{m,{n_r}}$ serves as a Gaussian-noise–corrupted approximation to ${Z_{m,{n_r}}}$ and $V_{m,{n_r}}^P $ is the effective noise variance. The process of our proposed GAMP-GESBL for channel estimation and target sensing scheme is summarized in \textbf{Algorithm \ref{GAMP-GE-SBL}}. ${\vartheta _s}$, ${\vartheta _h} \in (0,1]$ are the damping factors to moderate the update steps and promote stable convergence.
In the $q$-th iteration of GAMP, the input and output functions ${g_{\bar H}}\left( U_{j,{n_r}}^{\left( q \right)},V_{j,{n_r}}^{U\left( q \right)},\bm{\zeta}^{(g-1)} \right)$ and ${g_S}\left( P_{m,{n_r}}^{\left( q \right)},V_{m,{n_r}}^{P\left( q \right)},\bm{\zeta}^{(g-1)} \right)$ are constructed from the approximate posterior distributions of the $\bar H_{m,n_r}$ and $Z_{m,n_r}$, with the expression of 
\begin{equation}
    {g_{\bar H}}\left( U_{j,{n_r}}^{\left( q \right)},V_{j,{n_r}}^{U\left( q \right)},\bm{\zeta}^{(g-1)} \right)=\mu _{j,{n_r}}^{\bar H (q)} =\frac{{{U^{(q)}_{j,{n_r}}}}}{{{\delta^{(g-1)} _j}V_{j,{n_r}}^{U(q)} + 1}},
\end{equation}
and
\begin{equation}
\begin{split}
 &{g_S}\left( {P_{m,{n_r}}^{\left( q \right)},V_{m,{n_r}}^{P\left( q \right)},{\bm{\zeta}^{\left( g-1 \right)}}} \right) = \frac{{\mu _{m,{n_r}}^{Z\left( q \right)} - P_{m,{n_r}}^{\left( q \right)}}}{{V_{m,{n_r}}^{P\left( q \right)}}}\\
 &=\frac{1}{{V_{m,{n_r}}^{P\left( q \right)}}}\left( {\frac{{V_{m,{n_r}}^{P\left( q \right)}{\beta ^{\left( g-1 \right)}}{Y_{m,{n_r}}} + P_{m,{n_r}}^{\left( q \right)}}}{{{\beta ^{\left( g-1 \right)}}V_{m,{n_r}}^{P\left( q \right)} + 1}} - P_{m,{n_r}}^{\left( q \right)}} \right),
\end{split}
\end{equation}
respectively. More details about GAMP can be found in {\cite{GAMP6033942}}.

\begin{figure}
\centering
\includegraphics[width=8cm]{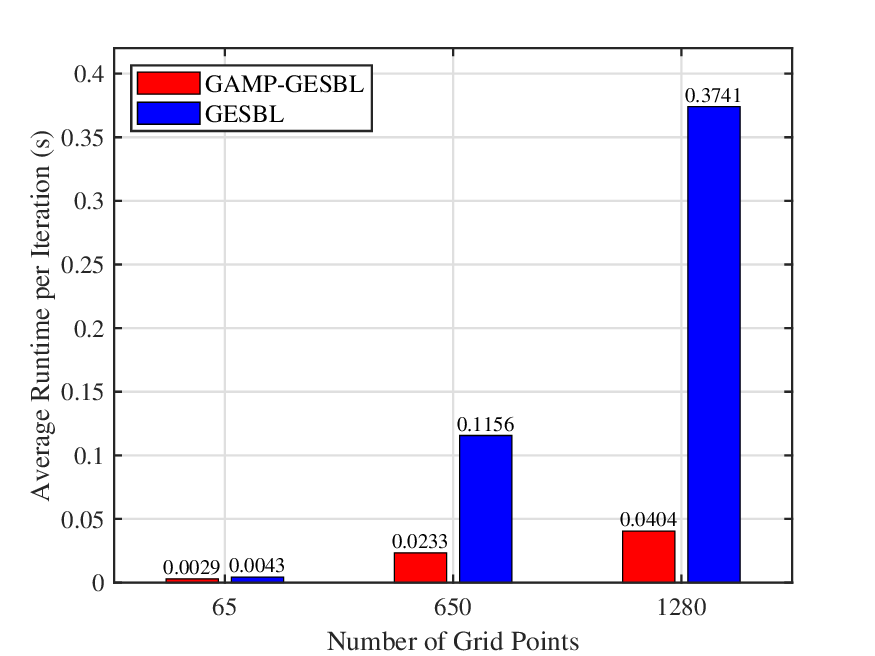}
\caption{{Comparison of the average runtime per iteration between the proposed GAMP-GESBL and GESBL channel estimation schemes.}}
\label{run_time}
\end{figure}

The complexity of the proposed GAMP-GESBL for data-aided framework in Algorithm \ref{GAMP-GE-SBL} is analyzed as follows. Similar to Algorithm \ref{GE-SBL}, the update of $\hat{\mathbf{\Phi }}\left( \bm{\bar \ell },\bm{\bar k } \right)$ in each iteration has the complexity of $\mathcal{O}\left(\bar RNL_\tau K_\nu\right)$. However, the estimation of $\bm{\mu}$ and $\bm{\Sigma}$ is inverse free and both have the linear complexity of $\mathcal{O}\left(QNL_\tau K_\nu N_r\right)$. The update of the hyperparameters is also same as Algorithm \ref{GE-SBL} with a total complexity of $\mathcal{O}\left(NL_\tau K_\nu N_r+NN_r+2(L_\tau K_\nu+\bar R^3+\bar R^2)\right)$. {Compared to GE-SBL, GAMP-GE-SBL avoids high-complexity matrix inversion and generally achieves more favorable computational complexity scaling, especially when the codebook is large. To provide a more intuitive comparison of the computational complexity, Fig.~\ref{run_time} compares the average runtime\footnote{{The average runtime is measured on Intel Core i7-10700 @ 2.90 GHz.}} per iteration of the two channel estimation schemes under different grid sizes. Here, we set the number of subcarriers $N=256$ and the number of receive antennas $N_r=8$ as an example.}

\section{Numerical Results}
This section presents numerical experiments that evaluate the effectiveness of the proposed D-GESBL and D-GAMP-GESBL schemes for the AFDM-ISAC system, where the results are obtained through Monte Carlo simulations. The AFDM–ISAC system operates at a carrier frequency of 60 GHz and employs $N=256$ subcarriers together with 4-QAM modulation. The base station is equipped with a ULA containing $N_r= 8$ antenna elements. We consider $R=3$ targets in this system and their AoAs are independently and randomly distributed from $\left[-\pi/2,\pi/2\right]$. The maximum normalized Doppler shift is considered to be $k_{max}=2$, which corresponds to a maximum speed of $540$ km/h. The maximum normalized delay is considered to be $\ell_{max}=12$, and the CPP length is accordingly set to $N_\text{cpp}=12$ to prevent inter-symbol interference (ISI). The chirp parameters used in the AFDM waveform are $c_1=\frac{9}{512}$ and $c_2=10^{-5}$. 
The normalized mean square error (NMSE) of $\mathbf{H}_{eff}$ is defined as 
\begin{equation}
    \text{NMSE}\left(\mathbf{H}_{eff}\right)=\frac{{\left\| {{{{\bf{\hat H}}}_{eff}} - {{\bf{H}}_{eff}}} \right\|_F^2}}{{\left\| {{{\bf{H}}_{eff}}} \right\|_F^2}}.
\end{equation}
\subsection{Genie Bound}
To better demonstrate the superiority of the proposed scheme in AFDM-ISAC systems, a genie bound is employed as a performance baseline. For the genie bound, it is assumed that the delay, Doppler shift, and data symbols are perfectly known. Based on \eqref{DAF-receive}, the estimation of $\mathbf{H}$ can be obtained by
\begin{equation}
    \mathbf{\hat H} = \left(\mathbf{\Psi}^H\mathbf{\Psi}+\sigma^2\mathbf{I}\right)^{-1}\mathbf{\Psi}^H\mathbf{Y},
\end{equation}
where $\mathbf{\Psi}=\mathbf{\Psi}_p+\mathbf{\Psi}_d$ and $\sigma^2$ is the noise variance. Then, the effective channel response in the DAF domain for the $n_r$-the receive antenna can be expressed as
\begin{equation}
    \mathbf{\hat H}_{eff,n_r}= \sum\limits_{i = 0}^{R} {\hat H}_{i,n_r}\mathbf{A}{{\mathbf{\Delta }}_{{\nu_i}}}{{\mathbf{\Pi }}^{\eta_i}}\mathbf{A}^H.
\end{equation}
where ${\hat H}_{i,n_r}$ is the $(i,n_r)$-th element of $\mathbf{\hat H}$.
\begin{figure}
\centering
\includegraphics[width=8cm]{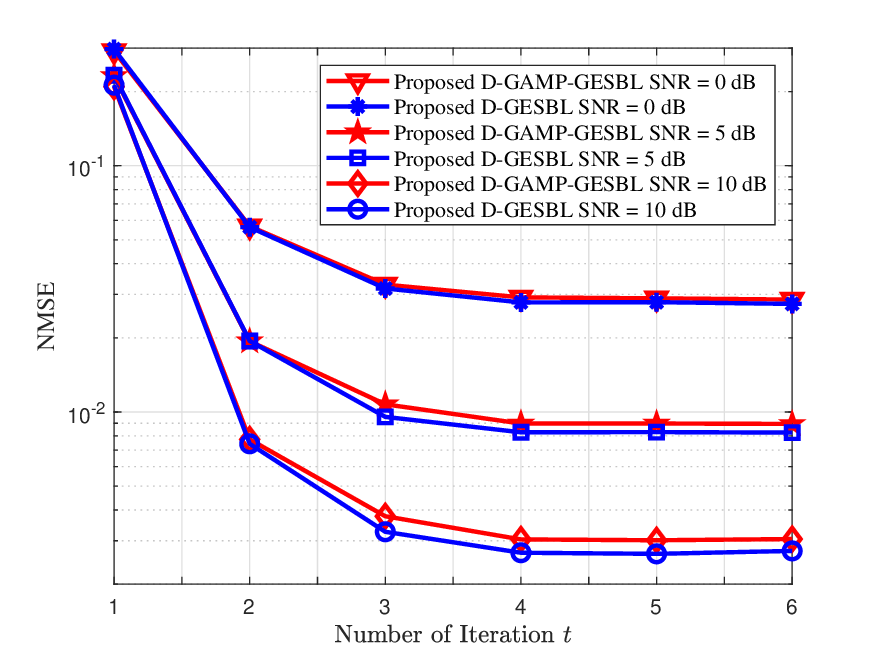}
\caption{The NMSE performance of $\mathbf{H}_{eff}$ estimation versus number of data-aided iteration $t$.}
\label{out_iteration}
\end{figure}

\begin{figure}
\centering
\includegraphics[width=8cm]{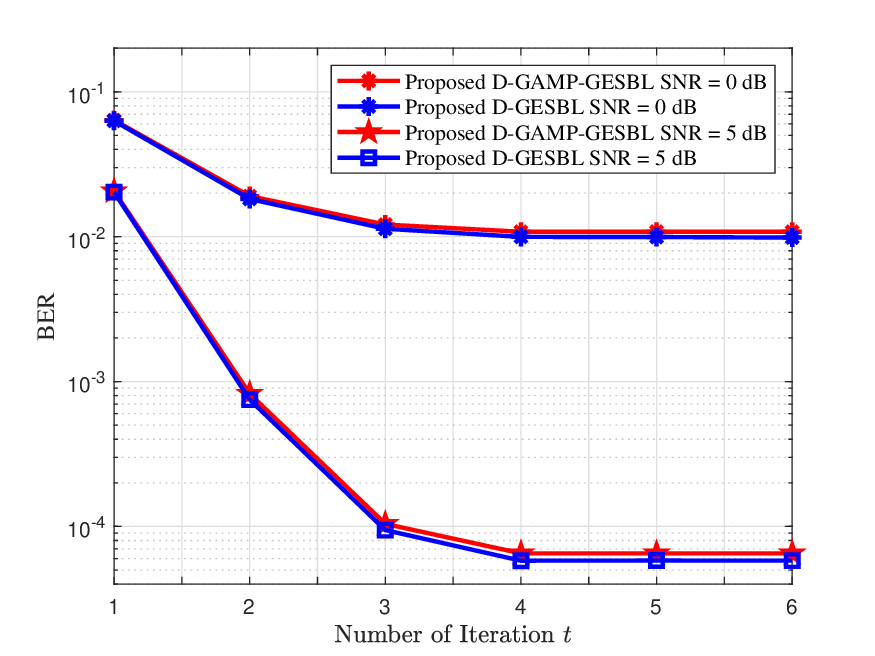}
\caption{{The BER performance versus number of data-aided iteration $t$.}}
\label{BER_iteration}
\end{figure}

\subsection{Simulation Results}
Fig.~\ref{out_iteration} illustrates the convergence behavior of the two proposed schemes under different SNR levels in terms of the channel estimation NMSE.
The resolutions of the virtual grids are set to be $r_\tau=1$ and $r_k=1$. As observed, the performance at $t=1$ is relatively poor because only the pilot symbols are available and the data symbols are treated as interference. Once the iterative process begins, the detected data progressively contributes to improving the sensing accuracy, while the refined channel sensing in turn enhances data detection. This mutually beneficial interaction drives the NMSE to decrease rapidly and reach a steady level after roughly four iterations, demonstrating the fast convergence of the proposed data-aided strategy. 
{This observation is further confirmed by the bit error rate (BER) evolution over the data-aided iterations in Fig.~\ref{BER_iteration}, where the representative SNR cases of $0$ dB and $5$ dB are shown to provide a clear illustration of the iterative BER improvement. As the iterative process proceeds, the BER decreases rapidly and gradually reaches a steady level, which further confirms the mutual enhancement between channel sensing and data detection in the proposed data-aided framework.}
Moreover, the proposed D-GAMP-GESBL scheme achieves a lower computational complexity than the proposed D-GESBL scheme without significantly performance loss.


\begin{figure}
\centering
\includegraphics[width=8cm]{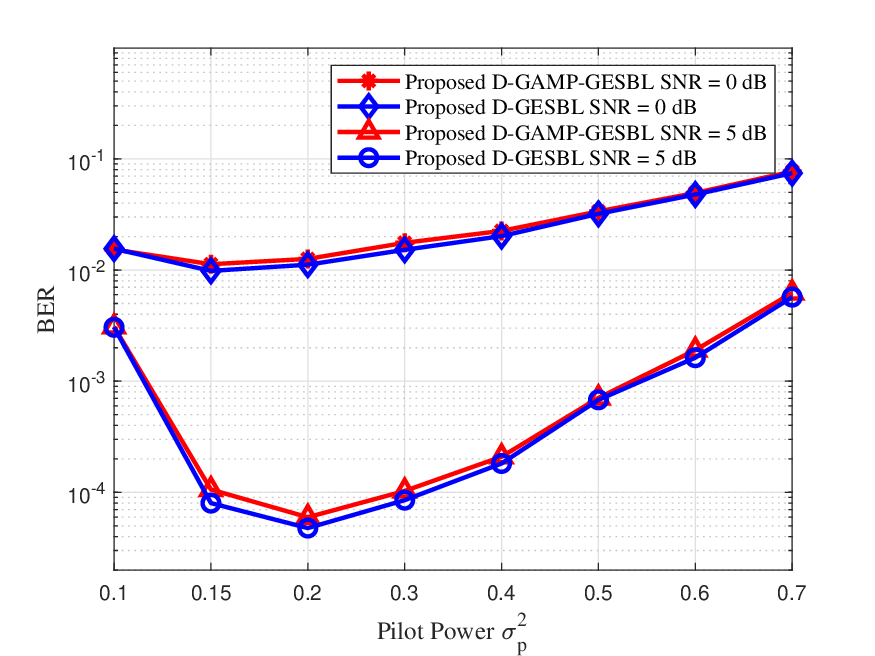}
\caption{{The BER performance versus pilot power $\sigma_p^2$.}}
\label{BER}
\end{figure}

Fig.~\ref{BER} illustrates the BER performance as a function of the pilot power under two SNR conditions, i.e., $\text{SNR}=0$ dB and $\text{SNR}=5$ dB, where the grid resolutions are set to $r_\tau=1$ and $r_k=1$. As observed, the two proposed schemes exhibit comparable performance, and all curves first improve and then degrade as the pilot power increases. The optimal pilot power occurs around $\sigma_p^2=0.2$. When the pilot power is less than the optimal value, insufficient pilot energy leads to poor channel sensing accuracy, resulting in degraded BER performance. Conversely, when the pilot power greater than the optimal value, although the channel sensing becomes more accurate, the reduced energy allocated to data transmission lowers the effective data SNR and eventually causes a deterioration in BER performance.

\begin{figure}
\centering
\includegraphics[width=8cm]{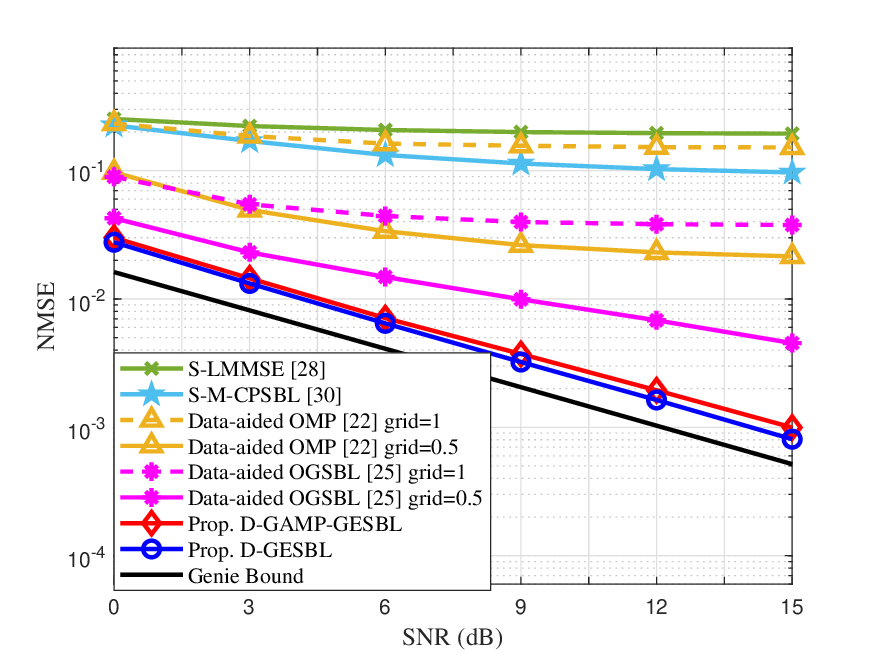}
\caption{The NMSE performance of $\mathbf{H}_{eff}$ estimation versus SNR compared with the state-of-the-art schemes.}
\label{compare}
\vspace{-0.2in}
\end{figure}

To better highlight the superiority of the proposed D-GESBL and D-GAMP-GESBL schemes for the AFDM-ISAC system, Fig.~\ref{compare} compares the estimation performance of the proposed D-GESBL and D-GAMP-GESBL schemes using $r_\tau=r_k=1$ with the existing state-of-the-art methods. Firstly, we compare with two SIC-based approaches, namely the SIC-based LMMSE (S-LMMSE) method in \cite{10711268} and the SIC-based multiple-frame coupled prior SBL (S-M-CPSBL) method in \cite{SP_SBL_SIC_10640141}. We also compare with two compressed sensing based methods, specifically the OMP algorithm in \cite{4DOMP9891774} and the OGSBL algorithm in \cite{OGSBL9738478}, both evaluated within the same data-aided framework as the proposed schemes using virtual grid resolutions of $r_\tau=r_k=0.5$ and $r_\tau=r_k=1$, which is shown as grid $=0.5$ and grid $=1$ in the figure, respectively. \textcolor{black}{Due to the reason that the SIC framework consistently regards the decoded data as interference rather than useful information and removes it, the effective sensing SNR is relative low. As a result, S-LMMSE and S-M-CPSBL cannot fully exploit the data-aided gain and therefore exhibit reduced NMSE performance compared with the proposed schemes. }
Within the data-aided framework, the performance of both OMP and OGSBL is highly sensitive to the resolution of the virtual grid. For the OMP algorithm, the mismatch between the true continuous parameters and the fixed grid introduces significant modeling errors. Although refining the grid resolution helps mitigate part of this mismatch, the overall estimation accuracy remains constrained. For OGSBL, the performance can be improved by estimating the off-grid components. However, the approximation error introduced by the first-order Taylor expansion cannot be eliminated. Employing a finer virtual grid reduces this error but comes at the cost of increased computational complexity. In contrast, our proposed schemes not only operate within the data-aided framework to leverage the decoded data into the codebook, but also dynamically refine the grid during the channel sensing process. By using the data-aided gain and iteratively adjusting the grids to better approximate the true parameters, the proposed schemes achieve significantly improved accuracy and closely approach the performance to that of the genie bound.

\begin{figure}
\centering
\includegraphics[width=8cm]{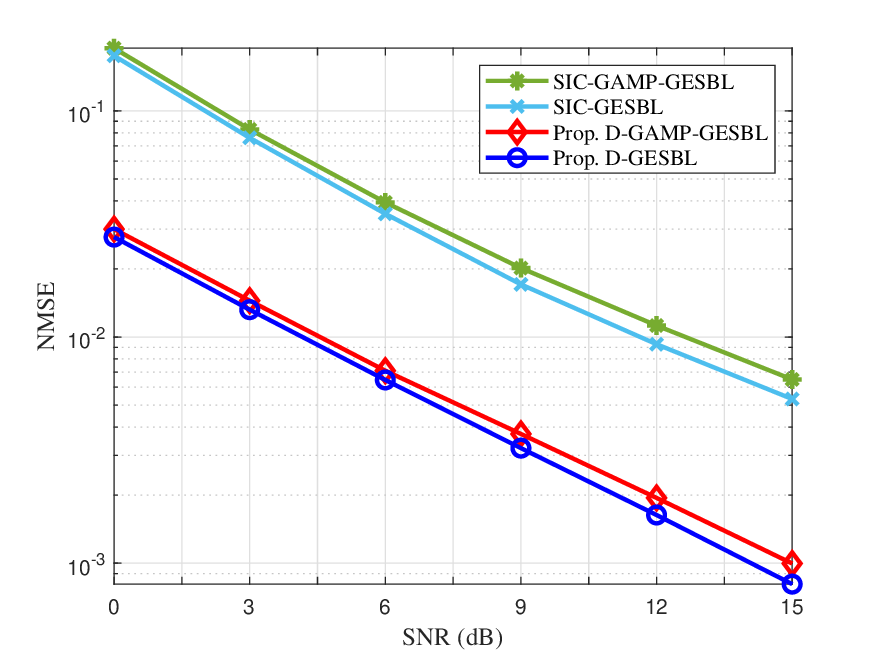}
\caption{The NMSE performance of $\mathbf{H}_{eff}$ estimation versus SNR.}
\label{SIC_data_aided}
\end{figure}

{To further demonstrate the superiority of the proposed data-aided strategy, Fig. \ref{SIC_data_aided} compares the NMSE performance of the proposed D-GESBL and D-GAMP-GESBL schemes with their SIC-based counterparts, where the SIC-based versions of GESBL and GAMP-GESBL are denoted as SIC-GESBL and SIC-GAMP-GESBL, respectively. As shown in Fig. \ref{SIC_data_aided}, all schemes achieve improved NMSE performance as the SNR increases. More importantly, the proposed D-GESBL and D-GAMP-GESBL schemes consistently outperform the corresponding SIC-based schemes over the entire SNR range. This improvement is attributed to the fact that the proposed data-aided schemes reuse reliably detected data symbols as additional known components in the codebook, thus providing more informative observations for channel estimation. In contrast, the SIC-based schemes only cancel the detected data symbols and do not further exploit their useful contribution to the estimation process, thereby limiting the effective sensing SNR and degrading the estimation performance.
}

\begin{figure}
\centering
\includegraphics[width=8cm]{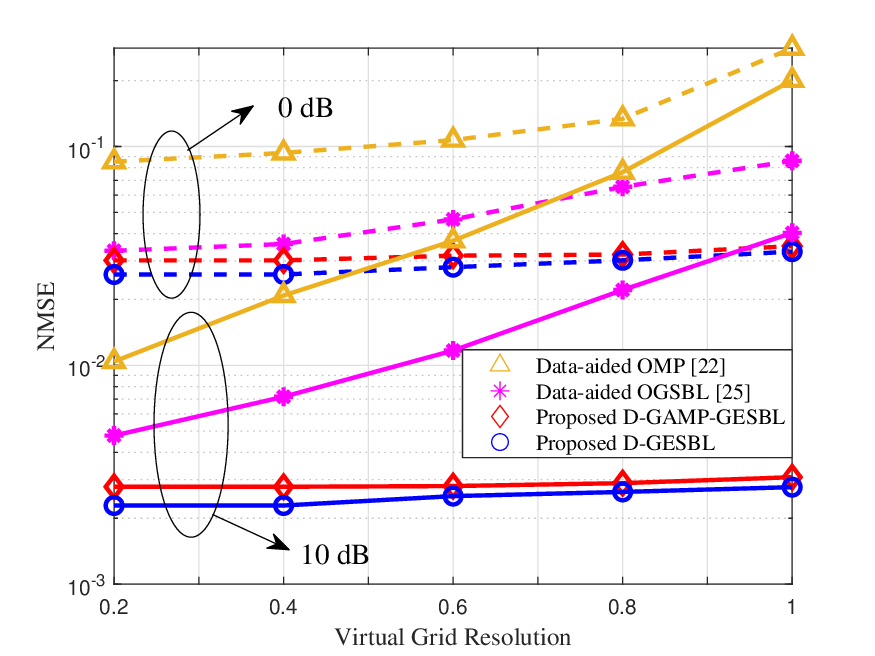}
\caption{The NMSE performance of $\mathbf{H}_{eff}$ estimation versus virtual grid resolutions ($r_\tau$, $r_k$).}
\label{grid_compare}
\end{figure}

The performance of the OMP method \cite{4DOMP9891774} and OGSBL method \cite{OGSBL9738478} is known to be sensitive to the resolution of the virtual grid. To clearly illustrate this effect, Fig.~\ref{grid_compare} compares these two baselines with our two proposed schemes under $\text{SNR}=0$ dB and $\text{SNR}=10$ dB for different virtual grid resolutions in data-aided framework. Here, we set $r_\tau=r_k$. For a fair comparison and to isolate the impact of the algorithms themselves, all sensing tasks are 
solved using a data-aided strategy. It can be observed that as the grid resolution becomes coarser, both OMP and OGSBL suffer noticeable NMSE degradation due to their reliance on a fixed virtual grid. Although OGSBL is able to account for off-grid components to some extent, it still fails to address the approximation error introduced by the first-order Taylor expansion. In contrast, the proposed schemes employ a dynamic virtual grid that iteratively refines the grid locations by estimating the off-grid offsets. This adaptive mechanism effectively mitigates the linear approximation error and steers the grid toward the true parameter values during the iterative process. As a result, the proposed methods exhibit negligible dependence on the initial grid resolution, meaning that desired performance can be achieved without resorting to finer grids that would otherwise increase computational complexity.

\begin{figure}
\centering
\includegraphics[width=8cm]{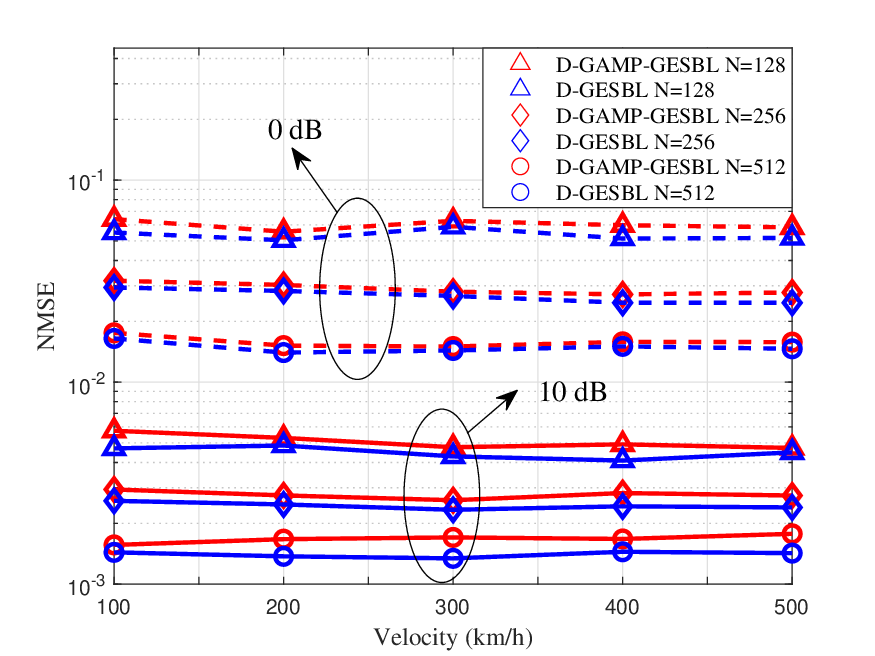}
\caption{The NMSE performance of $\mathbf{H}_{eff}$ estimation at different target velocities.}
\label{NMSE_vs_velocity}
\end{figure}

We further show the NMSE performance of the proposed two schemes for estimating $\mathbf{H}_{eff}$ across different numbers of subcarriers $N$ and various SNR conditions as the velocity of targets varies in Fig.~\ref{NMSE_vs_velocity}. It can be observed that all curves remain relatively stable over the entire velocity range, indicating that both algorithms exhibit strong robustness to high-mobility scenarios. Meanwhile, as the number of subcarriers increases, the delay resolution of the system is improved, leading to more accurate channel estimation performance.

\begin{figure}
\centering
\includegraphics[width=8cm]{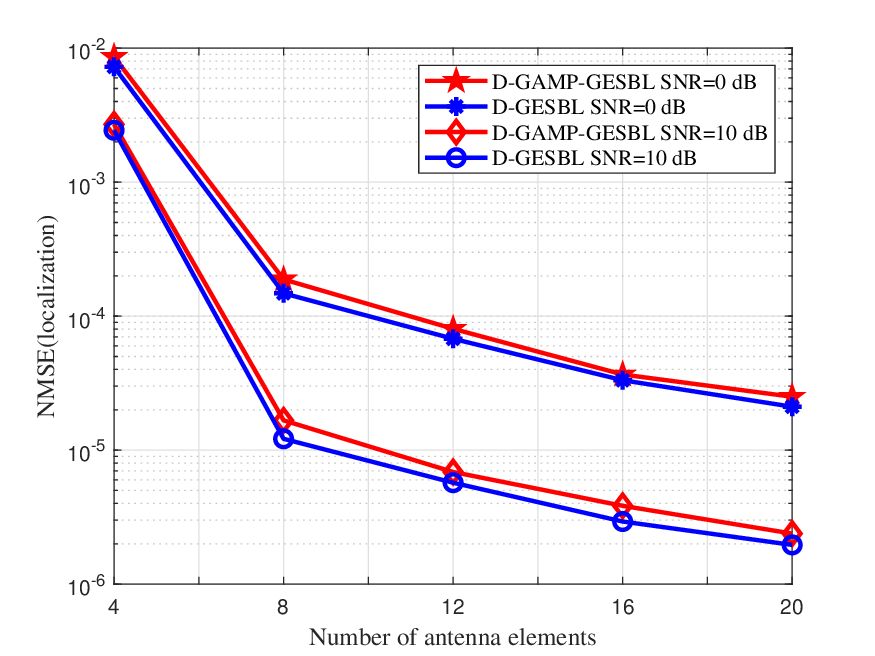}
\caption{The NMSE performance of localization as a function of the number of receive antenna elements.}
\label{localization}
\end{figure}

\begin{figure}
\centering
\includegraphics[width=8cm]{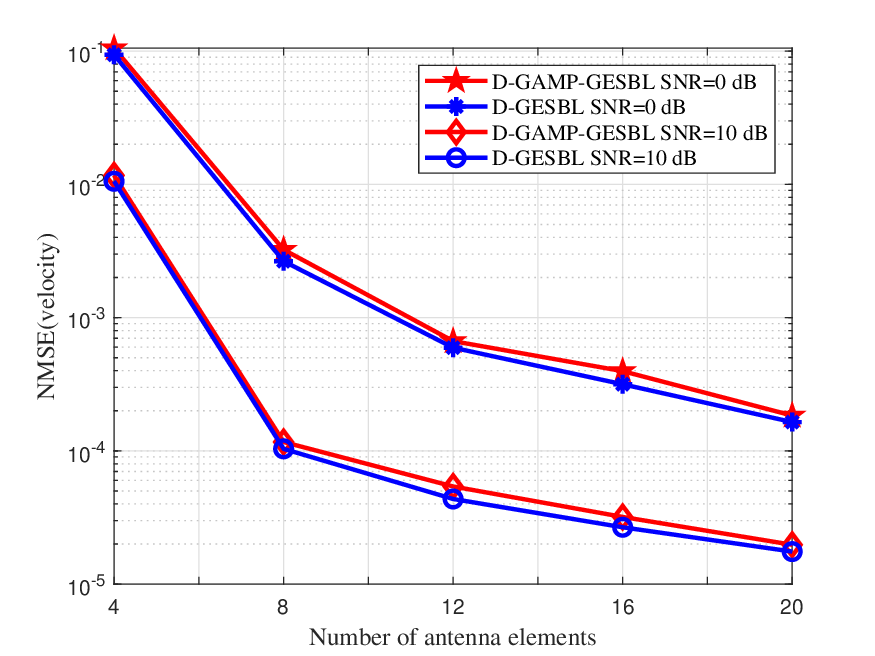}
\caption{The NMSE performance of velocity estimation as a function of the number of receive antenna elements.}
\label{velocity}
\end{figure}

Finally, Fig.~\ref{localization} and Fig.~\ref{velocity} show how the NMSEs of localization and velocity estimation vary with the number of receive antenna elements under $\text{SNR}=0$ dB and $\text{SNR}=10$ dB, respectively. In both figures, the curves follow a similar trend in which the estimation accuracy consistently improves as the number of antenna elements increase. This behavior results from the MMV formulation, where additional receive antennas contribute more independent observations and thus enhance the recovery performance. Moreover, for the localization, a larger number of receive antennas also leads to more accurate AoA estimation, which directly contributes to improved positioning accuracy.

\section{Conclusions}
In this paper, we proposed a D-GESBL channel estimation and sensing scheme for AFDM-ISAC systems. Specifically, to enhance the channel estimation and sensing accuracy under the superimposed pilot structure, we developed a data-aided strategy that exploits detected data symbols as additional pseudo-pilots information, while the overall channel estimation task is formulated as a MMV off-grid sparse recovery problem equipped with a grid evolution mechanism, which iteratively updates the virtual grids in the DAF domain to mitigate off-grid mismatch. Moreover, by integrating the GAMP algorithm into the SBL framework, we developed a D-GAMP–GESBL scheme that avoids matrix inversion and thereby achieves a substantial reduction in computational complexity without significantly performance loss. 
In addition, the numerical results verified the accuracy and robustness of our proposed two schemes and consistently showed their superiority over the existing benchmark methods.
\appendices
{\section{Proof of Proposition 1}}
To update $\bm{\delta}$, the objective function in \eqref{maxQ_delta} has the expression of \eqref{A_maxQ_delta} on the top of the next page.
\begin{figure*}[!t]
{\begin{equation}\label{A_maxQ_delta}
\begin{split}
    &Q\left( {{\bm{\delta }}|{{\bm{\zeta }}^{\left( {g - 1} \right)}},{{{\bm{\bar k}}}^{\left( {g - 1} \right)}},{{{\bm{\bar \ell}}}^{\left( {g - 1} \right)}}} \right)\\
    &={\mathbb{E}_{{\bf{\bar H}}|{\bf{Y}};{{\bm{\zeta }}^{\left( {g - 1} \right)}},{{{\bm{\bar k}}}^{\left( {g - 1} \right)}},{{{\bm{\bar \ell}}}^{\left( {g - 1} \right)}}}}\left\{ {\log \left( {p\left( {{\bf{\bar H}}|{\bm{\delta }}} \right)p\left( {\bm{\delta }} \right)} \right)} \right\}\\
    &={\mathbb{E}_{{\bf{\bar H}}|{\bf{Y}};{{\bm{\zeta }}^{\left( {g - 1} \right)}},{{{\bm{\bar k}}}^{\left( {g - 1} \right)}},{{{\bm{\bar \ell}}}^{\left( {g - 1} \right)}}}}\left\{ { - \sum\limits_{j = 0}^{{L_\tau }{K_\nu } - 1} {\left( {N_r\log {\delta _j} + b{\delta _j}} \right)}  - \sum\limits_{{n_r} = 0}^{{N_r} - 1} {\left( {{\bf{\bar H}}_{{n_r}}^H{{\bf{\Delta }}^{ - 1}}{{{\bf{\bar H}}}_{{n_r}}}} \right)} } +const\right\}\\
    &=- \sum\limits_{j = 0}^{{L_\tau }{K_\nu } - 1} {\left( {N_r\log {\delta _j} + b{\delta _j}} \right)} -{\mathbb{E}_{{\bf{\bar H}}|{\bf{Y}};{{\bm{\zeta }}^{\left( {g - 1} \right)}},{{{\bm{\bar k}}}^{\left( {g - 1} \right)}},{{{\bm{\bar \ell}}}^{\left( {g - 1} \right)}}}}\left\{ {\sum\limits_{{n_r} = 0}^{{N_r} - 1} {\left( {{\bf{\bar H}}_{{n_r}}^H{{\bf{\Delta }}^{ - 1}}{{{\bf{\bar H}}}_{{n_r}}}} \right)} } \right\}+const
\end{split}
\end{equation}
\noindent\rule{\textwidth}{0.6pt}
}  
\end{figure*}
As commonly adopted in the literature \cite{OGSBL9738478}, by setting the derivative of $Q\left( {{\bm{\delta }}|{{\bm{\zeta }}^{\left( {g - 1} \right)}},{{{\bm{\bar k}}}^{\left( {g - 1} \right)}},{{{\bm{\bar \ell}}}^{\left( {g - 1} \right)}}} \right)$ with respect to $\delta_j$ to zero, we obtain the solution to
\begin{equation}
\begin{split}
    &\frac{{\partial Q\left( {{\bm{\delta }}|{{\bm{\zeta }}^{\left( {g - 1} \right)}},{{{\bm{\bar k}}}^{\left( {g - 1} \right)}},{{{\bm{\bar \ell}}}^{\left( {g - 1} \right)}}} \right)}}{{\partial {\delta _j}}}\\
    &=\frac{{{N_r}}}{{{\delta _j}}} + b + \frac{{\sum\limits_{{n_r} = 0}^{{N_r} - 1} {\left( {{{\left| {\mu _{j,{n_r}}^{\left( g \right)}} \right|}^2} + \Sigma _{j,j}^{\left( g \right)}} \right)} }}{{\delta _j^2}}=0,
\end{split}
\end{equation}
which serves as the closed-form update rule for $\delta_j$, corresponding to the expression in \eqref{delta}.

\begin{figure*}
{\begin{equation}\label{A_maxQ_beta}
\begin{split}
    &Q\left( {\beta|{{\bm{\zeta }}^{\left( {g - 1} \right)}},{{{\bm{\bar k}}}^{\left( {g - 1} \right)}},{{{\bm{\bar \ell}}}^{\left( {g - 1} \right)}}} \right)\\
    &={\mathbb{E}_{{\bf{\bar H}}|{\bf{Y}};{{\bm{\zeta }}^{\left( {g - 1} \right)}},{{{\bm{\bar k}}}^{\left( {g - 1} \right)}},{{{\bm{\bar \ell}}}^{\left( {g - 1} \right)}}}}\left\{ {\log \left( {{\cal N}\left( {{\bf{Y}}|{{\bf{\Phi }}^{\left( {t,g - 1} \right)}}{\bf{\bar H}},{\beta ^{ - 1}}{\bf{I}}} \right)} \right) + \log p\left( \beta  \right)} \right\}\\
    &=\left( {d - 1 + {N_r}{L_\tau }{K_\nu }} \right)\log \beta  - \left( {e + {\mathbb{E}_{{\bf{\bar H}}|{\bf{Y}};{{\bm{\zeta }}^{\left( {g - 1} \right)}},{{{\bm{\bar \ell}}}^{\left( {g - 1} \right)}},{{{\bm{\bar k}}}^{\left( {g - 1} \right)}}}}\left\{ {\left\| {{\bf{Y}} - {{\bf{\Phi }}^{\left( {t,g - 1} \right)}}{\bf{\bar H}}} \right\|_F^2} \right\}} \right)\beta+const \\
\end{split}
\end{equation}
\noindent\rule{\textwidth}{0.6pt}}  
\end{figure*}

{Similar for the update of $\beta$, the objective function in \eqref{maxQ_beta} has the expression of \eqref{A_maxQ_beta} {on the top of the next page.} By setting the derivative of $Q\left( {\beta|{{\bm{\zeta }}^{\left( {g - 1} \right)}},{{{\bm{\bar k}}}^{\left( {g - 1} \right)}},{{{\bm{\bar \ell}}}^{\left( {g - 1} \right)}}} \right)$ with respect to $\beta$ to zero, we obtain the solution to
\begin{equation}
\begin{split}
    &\frac{{\partial Q\left( {\beta|{{\bm{\zeta }}^{\left( {g - 1} \right)}},{{{\bm{\bar k}}}^{\left( {g - 1} \right)}},{{{\bm{\bar \ell}}}^{\left( {g - 1} \right)}}} \right)}}{{\partial {\beta}}}\\
    &=\frac{{d - 1 + {N_r}{L_\tau }{K_\nu }}}{\beta } - \left( {e + \left\{ {\left\| {{\bf{Y}} - {{\bf{\Phi }}^{\left( {t,g - 1} \right)}}{\bf{\bar H}}} \right\|_F^2} \right\}} \right)=0,
\end{split}
\end{equation}
which serves as the closed-form update rule for $\beta$, corresponding to the expression in \eqref{beta}.
}
\section{Proof of Proposition 2}
\begin{figure*}[b]
\noindent\rule{\textwidth}{0.6pt}  
\begin{equation}\label{Express_kappa}
\begin{split}
   &{\mathbb{E}_{{\bf{\bar H}}|{\bf{Y}};{{\bm{\zeta }}^{\left( {g - 1} \right)}},{{{\bm{\bar k}}}^{\left( {g - 1} \right)}},{{{\bm{\bar \ell}}}^{\left( {g - 1} \right)}}}}\left\{ {\left\| {{\bf{Y}} - \tilde{\mathbf{\Phi }}^{\left(t\right)}\left( \bm{\bar k }^{\left(g-1\right)},\bm{\bar \ell }^{\left(g-1\right)},\bm{\kappa},{\bm{\iota }} \right){\bf{\bar H}}} \right\|_F^2} \right\}\\
   &=\sum\limits_{{n_r} = 0}^{{N_r} - 1} {\left\| {{{\bf{Y}}_{{n_r}}} - \tilde{\mathbf{\Phi }}^{\left(t\right)}\left( \bm{\bar k }^{\left(g-1\right)},\bm{\bar \ell }^{\left(g-1\right)},\bm{\kappa},{\bm{\iota }} \right)\mu _{{n_r}}^{\left( g \right)}} \right\|_2^2}+ {N_r}{\rm{Tr}}\left\{ {\tilde{\mathbf{\Phi }}^{\left(t\right)}\left( \bm{\bar k }^{\left(g-1\right)},\bm{\bar \ell }^{\left(g-1\right)},\bm{\kappa},{\bm{\iota }} \right){{\bf{\Sigma }}^{\left( g \right)}}{{\tilde{\mathbf{\Phi }}^{\left(t\right)}\left( \bm{\bar k }^{\left(g-1\right)},\bm{\bar \ell }^{\left(g-1\right)},\bm{\kappa},{\bm{\iota}} \right)}^H}} \right\}
\end{split}
\end{equation}
\vspace{-0.45in}
\end{figure*}
The expectation in \eqref{E_kappa} can be further expressed as \eqref{Express_kappa}, which is shown \textcolor{black}{on the bottom of the next page.} The first component of \eqref{Express_kappa} can be expressed as 
\begin{equation}\label{first term}
\begin{split}
    &\sum\limits_{{n_r} = 0}^{{N_r} - 1} {\left\| {{{\bf{Y}}_{{n_r}}} - \tilde{\mathbf{\Phi }}^{\left(t\right)}\left( \bm{\bar k }^{\left(g-1\right)},\bm{\bar \ell }^{\left(g-1\right)},\bm{\kappa},{\bm{\iota}} \right)\mu _{{n_r}}^{\left( g \right)}} \right\|_2^2}\\
    &=\sum\limits_{{n_r} = 0}^{{N_r} - 1} {{{\bm{\kappa }}^H}{\mathop{\rm Re}\nolimits} \left\{ {{{\left( {{\bf{\hat \Phi }}_{\bf{B}}^{\left( {t,g - 1} \right)H}{\bf{\hat \Phi }}_{\bf{B}}^{\left( {t,g - 1} \right)}} \right)}^ * } \circ \left( {\bm{\mu} _{{n_r}}^{\left( g \right)}\bm{\mu} _{{n_r}}^{\left( g \right)H}} \right)} \right\}{\bm{\kappa }}}\\
    &+\sum\limits_{{n_r} = 0}^{{N_r} - 1} {{{\bm{\iota }}^H}{\mathop{\rm Re}\nolimits} \left\{ {{{\left( {{\bf{\hat \Phi }}_{\bf{C}}^{\left( {t,g - 1} \right)H}{\bf{\hat \Phi }}_{\bf{C}}^{\left( {t,g - 1} \right)}} \right)}^ * } \circ \left( {\bm{\mu} _{{n_r}}^{\left( g \right)}\bm{\mu} _{{n_r}}^{\left( g \right)H}} \right)} \right\}{\bm{\iota }}}\\
    &+2\sum\limits_{{n_r} = 0}^{{N_r} - 1} { {{{\bm{\kappa }}^H}{\mathop{\rm Re}\nolimits} \left\{ {{{\left( {{\bf{\hat \Phi }}_{\bf{B}}^{\left( {t,g - 1} \right)H}{\bf{\hat \Phi }}_{\bf{C}}^{\left( {t,g - 1} \right)}} \right)}^ * } \circ \left( {\bm{\mu} _{{n_r}}^{\left( g \right)}\bm{\mu} _{{n_r}}^{\left( g \right)H}} \right)} \right\}{\bm{\iota }}}}\\
    &-2\sum\limits_{{n_r} = 0}^{{N_r} - 1} {{\mathop{\rm Re}\nolimits} \left\{ {{\bf{\tilde Y}}_{{n_r}}^{\left( {t,g} \right)H}{\bf{\hat \Phi }}_{\bf{B}}^{\left( t,g-1 \right)}{\rm{diag}}\left( {\bm{\mu} _{{n_r}}^{\left( g \right)}} \right){\bm{\kappa }}} \right\}}\\
    &-2\sum\limits_{{n_r} = 0}^{{N_r} - 1} {{\mathop{\rm Re}\nolimits} \left\{ {{\bf{\tilde Y}}_{{n_r}}^{\left( {t,g} \right)H}{\bf{\hat \Phi }}_{\bf{C}}^{\left( t,g-1 \right)}{\rm{diag}}\left( {\bm{\mu} _{{n_r}}^{\left( g \right)}} \right){\bm{\iota }}} \right\}} +\text{const}   
\end{split}
\end{equation}
where ${\bf{\tilde Y}}_{{n_r}}^{\left( {t,g} \right)}$$={{\bf{Y}}_{{n_r}}} - {{\bf{\hat \Phi }}^{\left( {t,g - 1} \right)}} \bm{\mu} _{{n_r}}^{\left( g \right)}$, and ${\bf{\hat \Phi }}^{\left( {t,g - 1} \right)}$, ${\bf{\hat \Phi }}_{\bf{B}}^{\left( {t,g - 1} \right)}$, and ${\bf{\hat \Phi }}_{\bf{C}}^{\left( {t,g - 1} \right)}$ are the simplified expression of ${\bf{\hat \Phi }}^{\left( {t} \right)}\left(\bm{\bar k}^{(g-1)},\bm{\bar \ell}^{(g-1)}\right)$, ${\bf{\hat \Phi }}_{\bf{B}}^{\left( {t} \right)}\left(\bm{\bar k}^{(g-1)},\bm{\bar \ell}^{(g-1)}\right)$, and ${\bf{\hat \Phi }}_{\bf{C}}^{\left( {t} \right)}\left(\bm{\bar k}^{(g-1)},\bm{\bar \ell}^{(g-1)}\right)$, respectively.
The second component of \eqref{Express_kappa} can be expressed as \eqref{second term}, which is shown on the top of the next page.
\begin{figure*}
    \begin{equation}\label{second term}
    \begin{split}
        &{N_r}{\rm{Tr}}\left\{ \tilde{\mathbf{\Phi }}^{\left(t\right)}\left( \bm{\bar k }^{\left(g-1\right)},\bm{\bar \ell }^{\left(g-1\right)},\bm{\kappa},{\bm{\iota}} \right){{\bf{\Sigma }}^{\left( g \right)}}{{\tilde{\mathbf{\Phi }}^{\left(t\right)}\left( \bm{\bar k }^{\left(g-1\right)},\bm{\bar \ell }^{\left(g-1\right)},\bm{\kappa},{\bm{\iota }} \right)}^H}\right\} \\
        &=2N_r{\mathop{\rm Re}\nolimits} {\left\{ {{\bf{\hat \Phi }}_{\bf{B}}^{\left( {t,g - 1} \right)H}{\bf{\hat \Phi }}^{\left( {t,g - 1} \right)}{{\bf{\Sigma }}^{\left( g \right)}}} \right\}^T}{\bm{\kappa}}+2N_r{\mathop{\rm Re}\nolimits} {\left\{ {{\bf{\hat \Phi }}_{\bf{C}}^{\left( {t,g - 1} \right)H}{\bf{\hat \Phi }}^{\left( {t,g - 1} \right)}{{\bf{\Sigma }}^{\left( g \right)}}} \right\}^T}{\bm{\iota}} + N_r{{\bm{\kappa }}^T}\left( {{{\left( {{\bf{\hat \Phi }}_{\bf{B}}^{\left( {t,g - 1} \right)H}{\bf{\hat \Phi }}_{\bf{B}}^{\left( {t,g - 1} \right)}} \right)}^ * } \circ {{\bf{\Sigma }}^{\left( g \right)}}} \right){\bm{\kappa }}\\
        &+N_r{{\bm{\iota }}^T}\left( {{{\left( {{\bf{\hat \Phi }}_{\bf{C}}^{\left( {t,g - 1} \right)H}{\bf{\hat \Phi }}_{\bf{C}}^{\left( {t,g - 1} \right)}} \right)}^ * } \circ {{\bf{\Sigma }}^{\left( g \right)}}} \right){\bm{\iota }}+2N_r{{\bm{\kappa }}^T}{\mathop{\rm Re}\nolimits} {\left\{ {{{\left( {{\bf{\hat \Phi }}_{\bf{B}}^{\left( {t,g - 1} \right)H}{\bf{\hat \Phi }}_{\bf{C}}^{\left( {t,g - 1} \right)}} \right)}^ * } \circ {{\bf{\Sigma }}^{\left( g \right)}}} \right\}{\bm{\iota }}}+\text{const},
    \end{split}
\end{equation}
\vspace{-0in}
\noindent\rule{\textwidth}{0.6pt}  
\end{figure*}

Since all elements in vector $\bm{\kappa}$ are real-valued, we have $\bm{\kappa}^H=\bm{\kappa}^T$. By substituting \eqref{first term} and \eqref{second term} into \eqref{Express_kappa} and setting the derivatives of \eqref{Express_kappa} with respect to $\bm{\kappa}$ and $\bm{\iota}$ to zero, respectively, two coupled stationarity conditions are obtained. The off-grid parameters are then updated in an alternating manner, where one parameter is optimized while keeping the other fixed.
Specifically, the update rule for $\bm{\kappa}$ is
\begin{equation}
    {{\bm{\kappa }}^{\left( g \right)}} = \mathop {\min }\limits_{{\bm{\kappa }} \in {{\left[ { - \frac{{{r_k }}}{2},\frac{{{r_k }}}{2}} \right]}^{{L_\tau }{K_\nu }}}} \left\{ {{{\bm{\kappa }}^T}{{\bf{Z}}^{\kappa \left( {t,g} \right)}}{\bm{\kappa }} - 2{{\bm{\alpha }}^{\kappa \left( {t,g} \right)}}{\bm{\kappa }}} \right\},
\end{equation}
and the closed form update rule is \eqref{kappa}. 
The update rule of $\bm{\iota}$ is similar and can be expressed as
\begin{equation}
    {{\bm{\iota }}^{\left( g \right)}} = \mathop {\min }\limits_{{\bm{\iota }} \in {{\left[ { - \frac{{{r_\tau }}}{2},\frac{{{r_\tau }}}{2}} \right]}^{{L_\tau }{K_\nu }}}} \left\{ {{{\bm{\iota }}^T}{{\bf{Z}}^{\iota \left( {t,g} \right)}}{\bm{\iota }} - 2{{\bm{\alpha }}^{\iota \left( {t,g} \right)}}{\bm{\iota }}} \right\},
\end{equation}
and the closed form update rule is \eqref{iota}. 
It is worth noting that, prior to each iteration, the off-grid components are initialized to zero. Consequently, during the update of $\bm{\kappa}$, the corresponding $\bm{\iota}^{(g-1)}$ remains zero, which allows the corresponding term to be neglected in the computation. However, to enhance the accuracy of the first-order Taylor approximation, the updated $\bm{\kappa}^{(g)}$ is subsequently utilized to assist in the update of $\bm{\iota}$. That's why \eqref{alpha_kappa} and \eqref{alpha_iota} have different expressions.

\bibliographystyle{IEEEtran}
\footnotesize
\bibliography{ref_AFDM}

\end{document}